%% file: main.tex
\documentclass[twocolumn]{aastex63}

\usepackage[utf8]{inputenc}
\usepackage{graphicx}
\graphicspath{{./}{./figures/}{./tex/figures/}}
\usepackage{amsmath}
\usepackage{afterpage}
\usepackage{enumitem}
\usepackage{xcolor}
\setenumerate{itemsep=0mm}
\usepackage{float}
\usepackage{soul} 
\usepackage{amsmath}
\usepackage{amssymb}
\usepackage{xspace}
\usepackage{xifthen}
\usepackage{eso-pic}
\usepackage{booktabs} 
\usepackage{verbatimbox} 
\usepackage{lineno} 

\newcommand{\vs}{vs.\xspace}
\newcommand{\roughly}{\ensuremath{ {\sim}\,} }
\newcommand{\var}[1]{\ensuremath{\texttt{\MakeUppercase{#1}}}\xspace}

\received{2021 July 8}
\revised{2021 August 30}
\accepted{2021 August 31}

\submitjournal{AJ}
\shorttitle{RR Lyrae stars in Centaurus I}
\shortauthors{Mart\'{i}nez-V\'{a}zquez et al.}

\graphicspath{{./}{figures/}}

\begin{document}
\reportnum{\footnotesize FERMILAB-PUB-21-274-AE-LDRD}

\title{RR Lyrae stars in the newly discovered ultra-faint dwarf galaxy Centaurus I\footnote{Based on DECam data}}

\correspondingauthor{Clara E. Mart\'{i}nez-V\'{a}zquez}
\email{clara.martinez@noirlab.edu}

\author[0000-0002-9144-7726]{C.~E.~Mart\'{i}nez-V\'azquez}
\affiliation{Cerro Tololo Inter-American Observatory, NSF's NOIRLab, Casilla 603, La Serena, Chile}
\author[0000-0003-1697-7062]{W.~Cerny}
\affiliation{Kavli Institute for Cosmological Physics, University of Chicago, Chicago, IL 60637, USA}
\affiliation{Department of Astronomy and Astrophysics, University of Chicago, Chicago IL 60637, USA}
\author[0000-0003-4341-6172]{A.~K.~Vivas}
\affiliation{Cerro Tololo Inter-American Observatory, NSF's NOIRLab, Casilla 603, La Serena, Chile}
\author[0000-0001-8251-933X]{A.~Drlica-Wagner}
\affiliation{Fermi National Accelerator Laboratory, P.O.\ Box 500, Batavia, IL 60510, USA}
\affiliation{Kavli Institute for Cosmological Physics, University of Chicago, Chicago, IL 60637, USA}
\affiliation{Department of Astronomy and Astrophysics, University of Chicago, Chicago IL 60637, USA}
\author[0000-0002-6021-8760]{A.~B.~Pace}
\affiliation{McWilliams Center for Cosmology, Carnegie Mellon University, 5000 Forbes Ave, Pittsburgh, PA 15213, USA}
\author{J.~D.~Simon}
\affiliation{Observatories of the Carnegie Institution for Science, 813 Santa Barbara St., Pasadena, CA 91101, USA}
\author{R.~R.~Munoz}
\affiliation{Departamento de Astronom\'ia, Universidad de Chile, Camino El Observatorio 1515, Las Condes, Santiago, Chile}
\author[0000-0002-7123-8943]{A.~R.~Walker}
\affiliation{Cerro Tololo Inter-American Observatory, NSF's NOIRLab, Casilla 603, La Serena, Chile}
\author{S.~Allam}
\affiliation{Fermi National Accelerator Laboratory, P.O.\ Box 500, Batavia, IL 60510, USA}
\author{D.~L.~Tucker}
\affiliation{Fermi National Accelerator Laboratory, P.O.\ Box 500, Batavia, IL 60510, USA}
\author[0000-0002-6904-359X]{M.~Adam\'ow}
\affiliation{Center for Astrophysical Surveys, National Center for Supercomputing Applications, 1205 West Clark St., Urbana, IL 61801, USA}
\author[0000-0002-3936-9628]{J.~L.~Carlin}
\affiliation{Rubin Observatory/AURA, 950 North Cherry Avenue, Tucson, AZ, 85719, USA}
\author{Y.~Choi}
\affiliation{Space Telescope Science Institute, 3700 San Martin Drive, Baltimore, MD 21218, USA}
\author[0000-0001-6957-1627]{P.~S.~Ferguson}
\affiliation{George P. and Cynthia Woods Mitchell Institute for Fundamental Physics and Astronomy, Texas A\&M University, College Station, TX 77843, USA}
\affiliation{Department of Physics and Astronomy, Texas A\&M University, College Station, TX 77843, USA}
\author{A.~P.~Ji}
\affiliation{Observatories of the Carnegie Institution for Science, 813 Santa Barbara St., Pasadena, CA 91101, USA}
\author{N.~Kuropatkin}
\affiliation{Fermi National Accelerator Laboratory, P.O.\ Box 500, Batavia, IL 60510, USA}
\author[0000-0002-9110-6163]{T.~S.~Li}
\affiliation{Observatories of the Carnegie Institution for Science, 813 Santa Barbara St., Pasadena, CA 91101, USA}
\affiliation{Department of Astrophysical Sciences, Princeton University, Princeton, NJ 08544, USA}
\affiliation{NHFP Einstein Fellow}
\author{D.~Mart\'{i}nez-Delgado}
\affiliation{Instituto de Astrof\'{i}sica de Andaluc\'{i}a, CSIC, E-18080 Granada, Spain}
\author[0000-0003-3519-4004]{S.~Mau}
\affiliation{Department of Physics, Stanford University, 382 Via Pueblo Mall, Stanford, CA 94305, USA}
\affiliation{Kavli Institute for Particle Astrophysics \& Cosmology, P.O.\ Box 2450, Stanford University, Stanford, CA 94305, USA}
\author[0000-0001-9649-4815]{B.~Mutlu-Pakdil}
\affiliation{Kavli Institute for Cosmological Physics, University of Chicago, Chicago, IL 60637, USA}
\affiliation{Department of Astronomy and Astrophysics, University of Chicago, Chicago IL 60637, USA}
\author{D.~L.~Nidever}
\affiliation{Department of Physics, Montana State University, P.O. Box 173840, Bozeman, MT 59717-3840; NSF's National Optical-Infrared Astronomy Research Laboratory, 950 N. Cherry Ave., Tucson, AZ 85719, USA}
\author{A.~H.~Riley}
\affiliation{George P. and Cynthia Woods Mitchell Institute for Fundamental Physics and Astronomy, Texas A\&M University, College Station, TX 77843, USA}
\affiliation{Department of Physics and Astronomy, Texas A\&M University, College Station, TX 77843, USA}
\author[0000-0002-1594-1466]{J.~D.~Sakowska}
\affiliation{Department of Physics, University of Surrey, Guildford GU2 7XH, UK}
\author[0000-0003-4102-380X]{D.~J.~Sand}
\affiliation{Department of Astronomy/Steward Observatory, 933 North Cherry Avenue, Room N204, Tucson, AZ 85721-0065, USA}
\author[0000-0003-1479-3059]{G.~S.~Stringfellow}
\affiliation{Center for Astrophysics and Space Astronomy, University of Colorado, 389 UCB, Boulder, CO 80309-0389, USA}

\collaboration{1000}{(DELVE Collaboration)}

\begin{abstract}
We report the detection of three RR Lyrae (RRL) stars (two RRc and one RRab) in the ultra-faint dwarf (UFD) galaxy Centaurus~I (Cen~I) and two Milky Way (MW) $\delta$~Scuti/SX~Phoenicis stars based on multi-epoch $giz$ DECam observations. The two RRc stars are located within 2 times the half-light radius (r$_h$) of Cen~I, while the RRab star (CenI-V3) is at $\sim6$\,r$_h$. The presence of three distant RRL stars clustered this tightly in space represents a 4.7$\sigma$ excess relative to the smooth distribution of RRL in the Galactic halo. Using the newly detected RRL stars, we obtain a distance modulus to Cen~I of $\mu_0 = 20.354 \pm 0.002$~mag ($\sigma=0.03$~mag), a heliocentric distance of D$_\odot = 117.7 \pm 0.1$~kpc ($\sigma=1.6$~kpc), with systematic errors of $0.07$~mag and $4$~kpc. The location of the Cen~I RRL stars in the Bailey diagram is in agreement with other UFD galaxies (mainly Oosterhoff~II). Finally, we study the relative rate of RRc+RRd (RRcd) stars ($f_{cd}$) in UFD and classical dwarf galaxies. The full sample of MW dwarf galaxies gives a mean of $f_{cd} = 0.28$. While several UFD galaxies, such as Cen~I, present higher RRcd ratios, if we combine the RRL populations of all UFD galaxies, the RRcd ratio is similar to the one obtained for the classical dwarfs ($f_{cd}$ $\sim$ 0.3). Therefore, there is no evidence for a different fraction of RRcd stars in UFD and classical dwarf galaxies.
\end{abstract}

\keywords{Dwarf galaxies (416), Stellar astronomy (1583), Local Group (929), Time domain astronomy (2109), Variable stars (1761), Pulsating variable stars (1307), RR Lyrae variable stars (1410)}

\section{Introduction} \label{sec:intro}

The $\Lambda$CDM cosmological model predicts that galaxies form hierarchically, with large galaxies formed by a continuous merging of low mass systems \citep{Searle1978,White1991,Frenk2012}. The dwarf satellite galaxies that we observe today may be the remnants of the merging process, and thus some authors refer to them as \textit{surviving representatives of the halo's building blocks} \citep[e.g.,][]{Fiorentino2015a}. The search for these building-blocks has provided the impetus for exceptional observational efforts targeting resolved Local Group dwarf systems \citep[e.g.,][]{Tolstoy2009}. However, the discovery of ultra-faint dwarf (UFD) galaxies located in the outer halo of our Galaxy has given a new perspective to the search for Galactic building blocks \citep[][and references therein]{Simon2019}. 
These numerous ($>$40), old ($>$10 Gyr), and metal-poor ([Fe/H] $< -2$ dex) systems can have extremely low present-day luminosities ($L \sim 10^3-10^5L_\odot$) and are considered to be among the most ancient relics of the formation of the Milky Way \citep[MW;][]{Bose2018}. 

Before the discovery of the first UFD galaxy a decade and a half ago \citep{Willman2005a, Willman2005b}, there was believed to be a clear distinction between dwarf galaxies and globular clusters, since they were situated in different locations in the absolute V-band magnitude (M$_V$) \vs physical half-light radius (r$_{1/2}$) plane (see, e.g. Figure 10 in \citealt{Willman2005a}). However, recently discovered systems with small sizes (r$_{1/2} \lesssim 80$ pc) and low luminosities (M$_V \gtrsim -6$ mag) cannot be definitively classified as star clusters or UFD galaxies without internal dynamics. Furthermore, the red giant branches of these systems are often so sparse, especially in shallow imaging, that their stellar populations and distances can only be determined at the most basic level. RR Lyrae (RRL) stars play an important role as unambiguous tracers of old stellar populations ($>$ 10 Gyr, \citealt{Walker1989, Savino2020}) and standard candles \citep[see, e.g.,][]{Beaton2018}. They are pulsating variable stars with periods between $\sim 7$ hours and $\sim 1$ day and with typical amplitudes of several tenths of a magnitude \citep{Smith1995, Catelan2015}. RRL stars are excellent primary distance indicators due to their well-established optical/near-infrared period-luminosity relations \citep[see e.g., ][]{Caceres&Catelan2008, Marconi2015}. Although the number of RRL stars in systems with M$_V > -3.5$~mag is expected to be of order 1$\pm$1 stars \citep[see Eq.~4 in][]{MartinezVazquez2019}, the detection of even a single RRL star offers an independent and accurate distance to the host system \citep[see e.g.,][]{Vivas2016a, MartinezVazquez2019}. Improving the distance measurement to a system allows a better determination of the physical size and absolute magnitude, thus helping to determine its nature. 

In addition, the period distribution of RRL stars provides clues about the contribution of the UFD galaxies to the formation of the MW halo \citep{Stetson2014, Zinn2014, Fiorentino2015a, Vivas2016a, Fiorentino2017}. While the inner halo has a period distribution peaked at P$\sim$0.55 days, the outer halo has a period distribution shifted to longer periods. Increasing the observed population of RRL stars in UFDs will help us to ascertain how much of the long-period tail of field halo RRL stars can be attributed to disrupted UFDs.

Centaurus~I (Cen~I) is an ultra-faint system (absolute magnitude M$_V = -5.5$~mag, azimuthally averaged half-light radius r$_h = 2.3\arcmin$) discovered by \citet{Mau2020} in the DECam Local Volume Exploration survey \citep[DELVE;][]{Drlica-Wagner2021}. DELVE combines archival DECam data with new observations to obtain complete coverage of the southern sky ($|b| > 10 \arcdeg$).  Data collection began in 2019A, with public DECam community data available through the NOIRLab Astro Data Archive\footnote{\url{https://astroarchive.noao.edu/}}.
Cen~I's measured age ($\tau > 12.85$~Gyr), size (r$_{1/2} = 79^{+14}_{-10} $ pc), and systemic metallicity ($\rm{[Fe/H]} = -1.8$~dex) place it within the size-magnitude locus consistent with most known UFDs \citep{Mau2020}.
UFDs with similar brightness as Cen~I have between 1 and 12 RRL stars \citep[see][]{MartinezVazquez2019, Vivas2020b}, and thus, we expect to detect several RRL in Cen~I. In fact, using the $N_{RRL}$ \vs $M_V$ relation from \citet[][equation 4]{MartinezVazquez2019}, Cen~I is expected to contain $6\pm 2$ RRL stars, which strongly motivates  high-cadence observations of this system.

The paper is structured as follows. In \S~\ref{sec:observations} we present our observations and explain the details of the data reduction process. In \S~\ref{sec:search}, we describe the search method we used for detecting variable stars in the field of Cen~I, and we report the variable stars detected in this work. In \S~\ref{sec:variable}, we present the classification of the variable stars detected, their light curves, and mean properties. We also show the color-magnitude diagram (CMD), the spatial distribution, and the proper motions (when available from \textit{Gaia}) of stars in Cen~I. In \S~\ref{sec:distance}, we determine the distances of the RRL stars associated with Cen~I. In \S~\ref{sec:oosterhoff}, we perform the Oosterhoff classification \citep{Oosterhoff1939, Oosterhoff1944} of Cen~I. In \S~\ref{sec:rrc}, we study the ratio of first overtone RRL stars in classical dwarf galaxies and UFDs. We investigate the angular size of Cen~I in \S~\ref{sec:extension}, and we conclude in \S~\ref{sec:conclusions}.

\section{Observations and Data} \label{sec:observations}
The data for this work were collected using the Dark Energy Camera (DECam, \citealt{Flaugher2015}) on the 4m Blanco Telescope located at the NSF's NOIRLab Cerro Tololo Inter-American Observatory (CTIO) in Chile. We obtained $g,i,z$ time-series photometry. 
The data were obtained in the second halves of 8--10 February 2020, 4--7 March 2020, and 15--19 March 2020 (PropID: 2020A-0238, P.I. Mart\'inez-V\'azquez).
We observed two dithered fields, one centering Cen~I on CCD N4 (one of the central DECam CCDs) and the other one dithering $60\arcsec$~ in RA and $60\arcsec$~ in Dec. from the previous pointing in order to cover the gaps between CCDs. 
The majority of the data were obtained in grey nights; however, we also used bright nights of director's discretionary time (where only $z,i$ observations were made). The mean (median) seeing of the images is $1.05\arcsec$ ($1.01\arcsec$) in $g$, $0.93\arcsec$ ($0.86\arcsec$) in $i$, and $0.91\arcsec$ ($0.87\arcsec$) in $z$. In total, we collected 98 exposures: 35 $g$-band, 39 $i$-band, and 25 $z$-band. Individual exposure times were 180s in $g$ and $i$, and 300s in $z$, which allowed us to reach $g,i,z$ $\sim 21$~mag with a $S/N \ga 50$ for single exposures.

We processed all exposures using the Dark Energy Survey (DES) Data Management (DESDM) pipeline \citep{Morganson:2018} following the procedure described in \citet{Drlica-Wagner2021}. The DESDM pipeline achieves sub-percent-level photometric accuracy by calibrating exposures based on seasonally--averaged bias and flat images and by performing full-exposure sky background subtraction \citep{Bernstein:2018}. This pipeline utilizes \verb|SourceExtractor| and \verb|PSFEx| \citep{Bertin:1996,Bertin:2011} for automatic source detection and photometric measurement on an exposure-level basis. Stellar positions are calibrated against \textit{Gaia} \citep{GaiaMission} Data Release 2  \citep[DR2;][]{GaiaDR2}, which provides 30 mas astrometric calibration precision. The photometry is calibrated by matching stars in each CCD to the ATLAS Refcat2 catalogs \citep{Tonry:2018}, which consists of measurements from Pan-STARRS DR1 (PS1; \citealt{Chambers:2016}) and SkyMapper DR1 \citep{Wolf:2018} transformed to the PS1 $griz$ filter system. For this calibration, stars were defined as objects passing a filter of $|\var{spread\_model\_(BAND)}| < 0.01$. Photometric measurements from the ATLAS Refcat2 catalog were transformed to the DECam $giz$ filters before calibration using the following empirically-derived equations:
{\par\small\begin{align*}
g_{\rm DECam} &= g_{\rm PS1} + 0.0994(g_{\rm PS1}-r_{\rm PS1}) - 0.0319 \\
i_{\rm DECam} &= i_{\rm PS1} -0.3407(i_{\rm PS1}-z_{\rm PS1}) - 0.0013 \\
z_{\rm DECam} &= r_{\rm PS1} -0.2575(r_{\rm PS1}-z_{\rm PS1}) - 0.0201,
\end{align*}\normalsize}
\noindent which have statistical root-mean-square (rms) calibration errors per CCD estimated relative to DES of $\roughly 0.01$ mag (see \citealt{Drlica-Wagner2021}). The typical photometric uncertainties for the HB stars of Cen~I are of the order of 0.005 mag.

In addition to our high-cadence observations, we also included individual $giz$ DECam images previously processed by DELVE in the field of Cen~I.\footnote{From March, 2013 and March-April, 2017.} These exposures were processed identically through the same pipeline described above. Further information on the DELVE photometric pipeline can be found in the \citet{Drlica-Wagner2021}.

\begin{figure*}
    \hspace{-1cm}
    \includegraphics[width=1.1\textwidth]{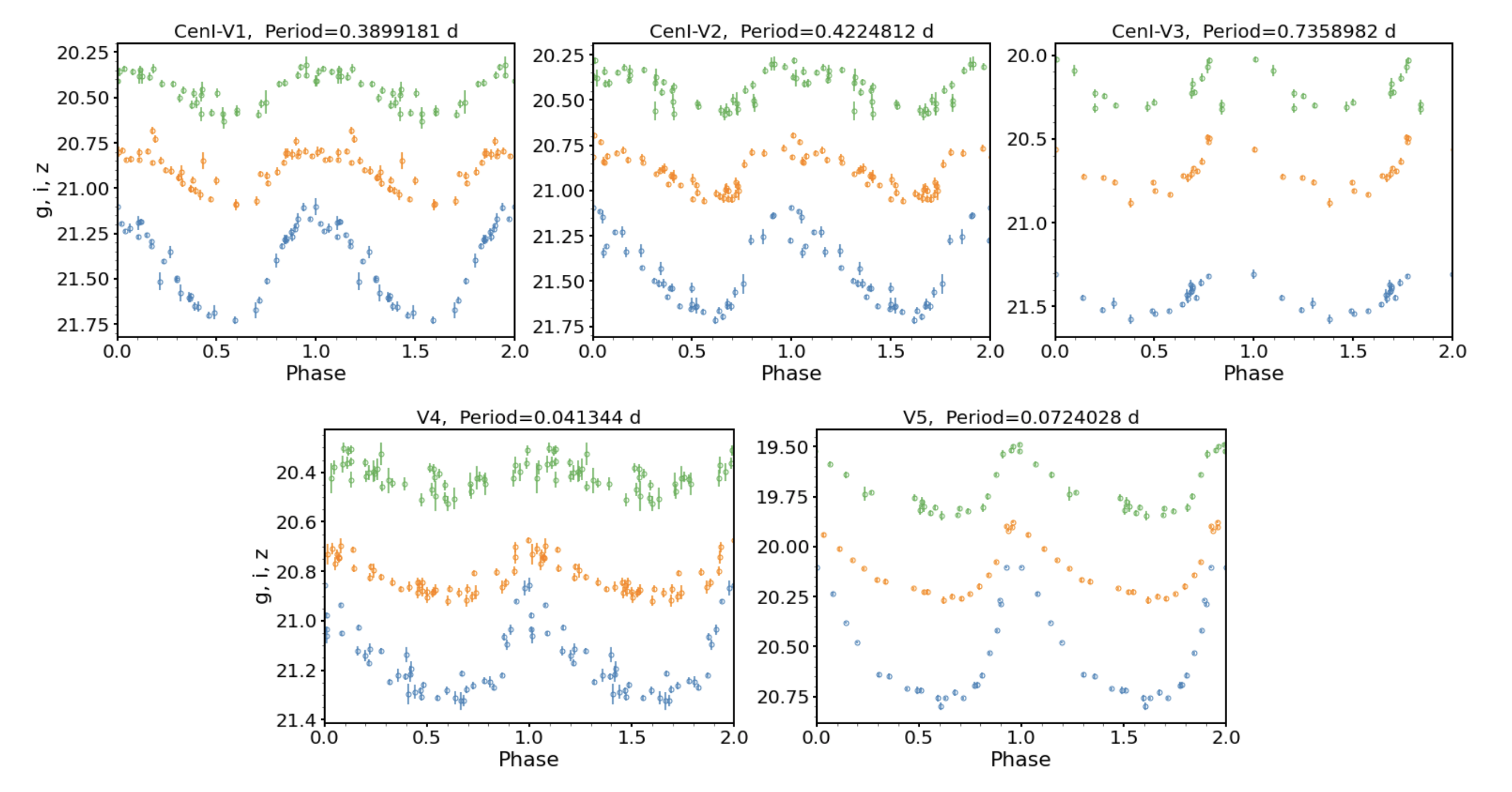}
    \caption{Light curves of the variable stars detected in the field of Cen I in the $g$ (blue), $i$ (orange) and $z$ (green) bands, phased with the period in days given at the top of each panel. The name of the variable is also displayed. For clarity, the $g$ and $z$ light curves have been shifted to $+0.2$ and $-0.4$ mag, respectively. RRL stars of Cen~I are displayed in the top panels while the field $\delta$~Sct/SX~Phe stars are in the bottom panels.}
    \label{fig:lcv}
\end{figure*}

\section{Search for variable stars}\label{sec:search}
To search for variable stars, we first constructed a multi-band source catalog by matching detections between individual exposures, following the procedure outlined in \citet{Drlica-Wagner2021}. For this catalog only, we cross-matched all unique sources detected in the individual exposures with a $0.7\arcsec$ matching radius, and calculated weighted-average photometry for each source based on their measurements in each exposure and their associated uncertainties. 

We calculated the extinction due to foreground dust from the MW for each individual source in the time-series and multi-band catalogs through bilinear interpolation from the rescaled versions of the extinction maps of \citet{Schlegel:1998} presented in \citet{Schlafly:2011}. We then calculated the reddening for each source by assuming a reddening law of $R_V = 3.1$ and utilizing a set of coefficients $R_{\lambda} = A_{\lambda}/E(B-V)$ derived by the Dark Energy Survey \citep{DR1:2018} for the $giz$ bands. 

We performed a search for periodic variable sources within $25\arcmin$ centered at the previously-identified centroid for Cen~I ($\alpha_{\rm{J2000}} = 189.585$ deg, $\delta_{\rm{J2000}} = -40.902$ deg) in the region of the CMD defined by $-0.5 \le (g-i) \le 0.6$ mag and $ 18.0 \le g \le 22.5$ mag. This region covers the instability strip of Cen~I, where pulsating variable stars are located --- specifically, RRL stars and Anomalous Cepheids (see e.g., Sculptor, \citealt{MartinezVazquez2016b}; or Sextans, \citealt{Vivas2019}). These selections produced hundreds of sources. As a variability index, we calculated a reduced chi-squared $\chi_{\nu}^2$ (see \citealt{Sok:2016}) for each star by comparing a given star individual PSF measurements to the median magnitude of that same star across all epochs for the $g$ band. Sources with $\chi_{\nu}^2 > 1$ were considered as potential variable candidates. We looked through all the time series in the sources selected to check whether they showed a reliable variation in their light curves. We produced periodograms as a Fourier transform of the time-series data following the prescription described in \citet{Horne1986}. The periodograms were calculated between 0.01 and 10 days, in order to encompass all the possible periods of RRL stars, Anomalous Cepheids and possible short periods variables, such as $\delta$~Scuti ($\delta$~Sct) or SX~Phoenicis (SX~Phe) stars. Once periodicity was confirmed, we obtained the first estimation of the period from the highest peak in the periodograms, but the final period was refined by visually inspecting the folded light curves in the three bands simultaneously. In addition, we visually inspected all the potential candidates in the images to remove spurious detections. The vast majority of spurious detections were background galaxies. Finally, we detected 3 RRL stars and 2 $\delta$~Sct/SX~Phe stars in our sample.

\section{Variable stars in the field of Cen~I} \label{sec:variable}

The most common types of RRL stars are the ab-type (RRab) and c-type (RRc). RRab stars are fundamental pulsators characterized by longer periods ($\sim$0.45--1.0~d) and saw-tooth light curves while RRc stars are first overtone pulsators with shorter periods ($\sim$0.2--0.45~d), lower amplitudes ($\Delta V \simeq 0.4$~mag), and almost sinusoidal light variations. We detected three RRL stars (2 RRc and 1 RRab) in the field of Cen~I. Since our photometry reaches several magnitudes below the HB and the observing strategy (cadence) was meant to search for RRL stars, we expect $\sim$100\% completeness for detecting isolated RRL stars.\footnote{We note that a variable source located within $0.7\arcsec$ of another source would be harder to recover due to the angular cross-matching that is performed to associate sources across individual exposures (see \S~\ref{sec:search})}

Assuming a smooth distribution of Galactic halo RRL stars \citep[see e.g.,][]{Vivas2016a, Zinn2014}, it is unlikely to find three RRL stars clumped together in space at large galactocentric distance. If we integrate the number density profile of RRL stars derived in \citet{Medina2018} --- which is appropriate for the outer Galactic halo out to distances of $\sim150$ kpc --- we find that 0.15 RRL stars are expected in a search area of 0.54 sq. deg. in the range of distances between 40 and 245 kpc (i.e., covering the magnitude limits of our search). The probability of finding three or more Galactic halo RRL stars within this region is $p = 5\times 10^{-4}$, which corresponds to a one-sided Gaussian significance of 3.3$\sigma$. In particular, if we estimate the number of MW halo RRL stars between 100 and 140 kpc that can contaminate our HB, the number is reduced to 0.02 RRL stars. The probability of finding three or more MW halo RRL stars in this case is $p = 1.3\times 10^{-6}$, which corresponds to a one-sided Gaussian significance of 4.7$\sigma$. Thus, these three RRLs are high confidence members of Cen~I. 

Additionally, we detected two $\delta$~Sct/SX~Phe variables of the MW. They are classified as $\delta$~Sct/SX~Phe stars because their periods are shorter than 0.1 d and their light curves are typical for this type of variable stars (see, e.g. \citealt{MartinezVazquez2021}). Furthermore, they are identified as MW field stars and not as members of Cen~I because they are pulsating main-sequence stars ($\delta$~Sct) or blue straggler stars (SX~Phe) and are thus significantly closer than Cen~I (see \citealt{Catelan2015}).

Figure~\ref{fig:lcv} shows the light curves in the different filters and Table~\ref{tab:photometry} provides the individual epoch photometry for all these variable stars. It is worth noting that the light curve of CenI-V3 has half the number of epochs (since it fell in one of the gaps between the DECam CCDs) and its phase space is not fully covered, particularly reflected as a lack of maximum in the light curve (see third top panel in Figure~\ref{fig:lcv}). We derive the pulsation parameters for the variable stars, obtaining the intensity-averaged magnitudes and amplitudes by fitting the light curves with the set of templates based on \citet{Layden1998}. The mean magnitudes were calculated using the best-fitting template, thus preventing biases in case light curves are not uniformly sampled in phase. Table~\ref{tab:rrl_prop} lists the coordinates and the pulsation parameters of the variable stars detected in the vicinity of Cen~I.

\input{tables/photometry_variables_CenI_tab_red}

\begin{verbbox}[\footnotesize]
ugali
\end{verbbox}

\begin{figure}
    \hspace{-0.5cm}
    \includegraphics[width=1.0\columnwidth]{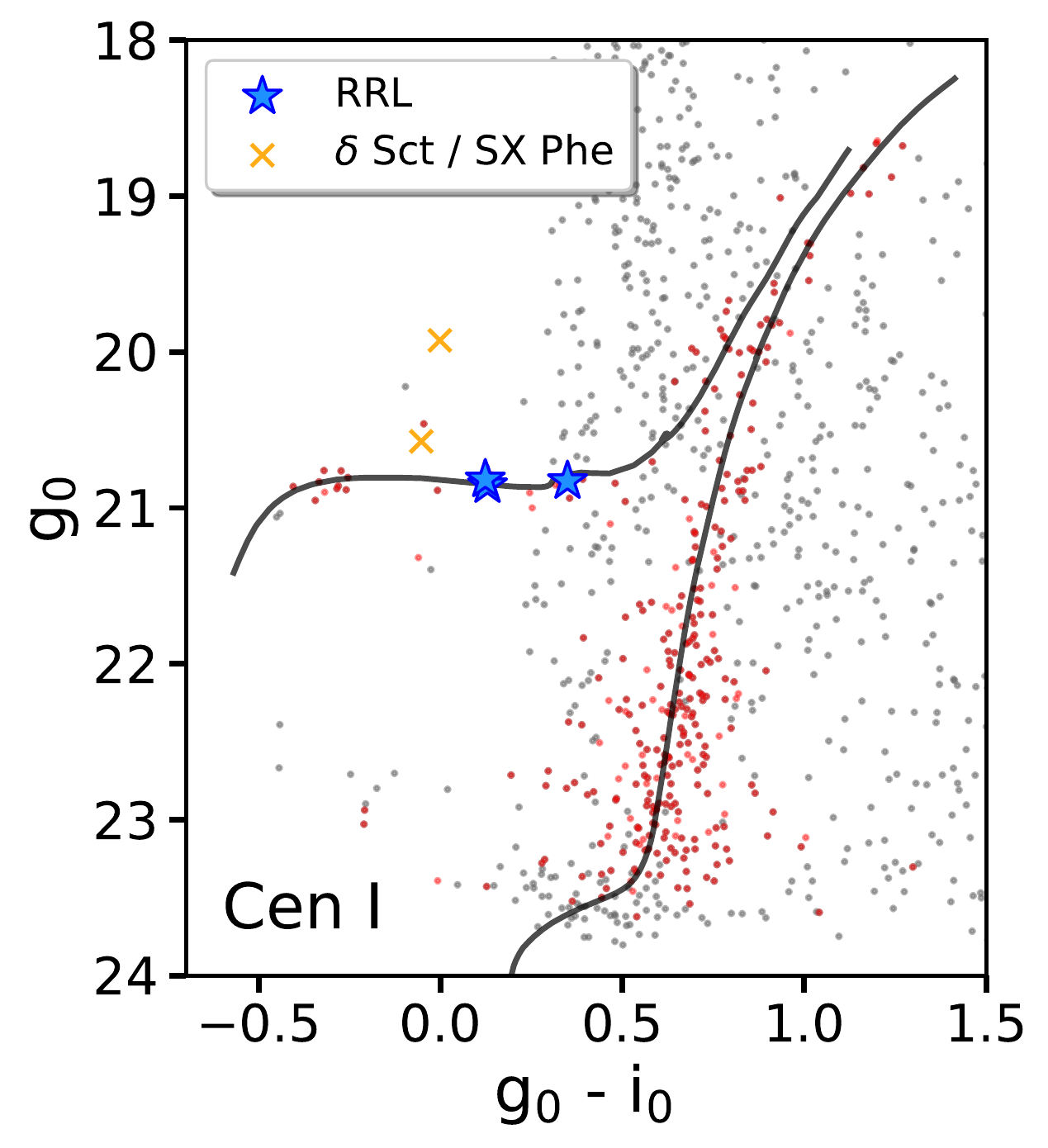}
    \caption{Dereddened CMD of Cen~I within $7\arcmin$ from the Cen~I center (grey points) and its three newly discovered RRL stars (blue stars). The black line is the isochrone of 12~Gyr and $Z=0.0001$ from BaSTI \citep{Hidalgo2018} shifted to a distance modulus of 20.35 mag (the distance modulus of Cen~I obtained in this work, \S~\ref{sec:distance}). The probable members of Cen~I ($p_{ugali} > 0.05$, \citealt{Mau2020}) are highlighted in red. Orange crosses are the MW field $\delta$~Sct/SX~Phe stars.}
    \label{fig:cmd}
\end{figure}

Figure~\ref{fig:cmd} displays the CMD of the stars found in the central $7\arcmin$ radius (grey points) of Cen~I, the candidate members of Cen~I according to \citet{Mau2020} (i.e., $p_{ugali} > 0.05$, red points),\footnote{This membership is based on the spatial position, measured flux, photometric uncertainty, and the local imaging depth, with an initial mass function weighting. It was obtained from the ultra-faint galaxy likelihood toolkit, \theverbbox: \url{https://github.com/DarkEnergySurvey/ugali} \citep{Bechtol2015, DrlicaWagner2015}.} and the variable stars found in this work. The three RRL stars (blue stars) are well positioned over the HB, while the two variables brighter than the HB (orange crosses) have periods shorter than 0.08 days and are field $\delta$~Sct/SX~Phe foreground stars.

\begin{figure}
    \hspace{-1.1cm}
    \includegraphics[width=1.2\columnwidth]{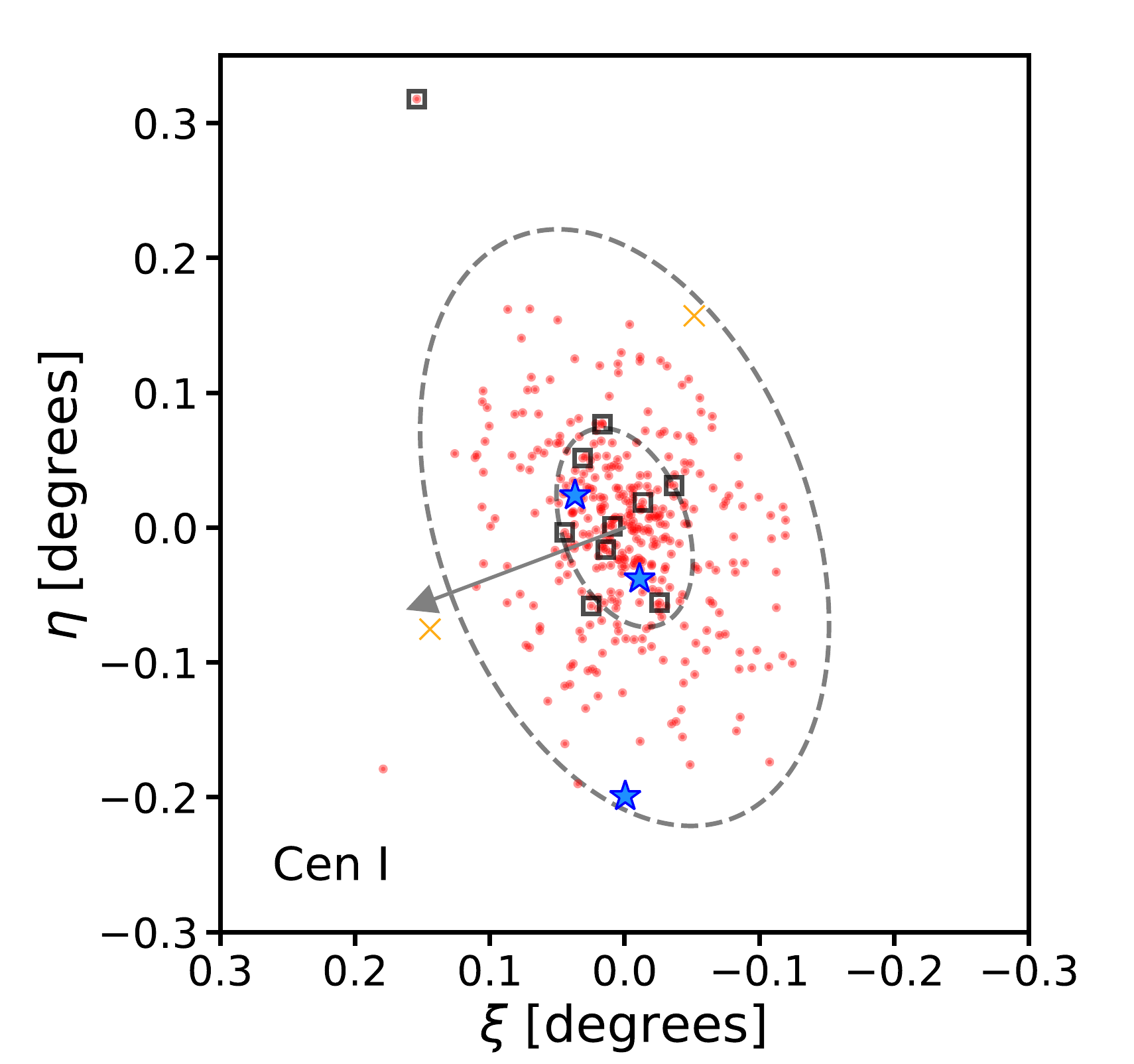}
    \caption{Spatial distribution in planar coordinates for the member candidates of Cen~I and the variable stars detected in the field. The three discovered RRL stars in Cen~I are shown as blue stars. The members of Cen~I are highlighted in red. The 10 BHB members of Cen~I are indicated by empty black squares. The ellipses correspond to 2 and 6 r$_h$ \citep[r$_h = 2.3\arcmin$,][]{Mau2020}. Orange crosses are MW field $\delta$~Sct/SX~Phe stars. The arrow marks the direction of the  reflex-corrected proper motion of Cen~I.}
    \label{fig:spatial}
\end{figure}

Figure~\ref{fig:spatial} shows the spatial distribution of the variable stars in the sky. Two of the RRL stars are within 2~r$_h$ (specifically between 1 and 2~r$_h$) while the third RRL is at $\sim$ 6~r$_h$. An examination of whether the latter is an extra-tidal member of Cen~I is presented \S~\ref{sec:extension}. 

\input{tables/rrl_prop}

\subsection{Cross-matching with \textit{Gaia} DR2 and \textit{Gaia} EDR3}\label{sec:gaia}

Within a radius of $25\arcmin$ (i.e., $\sim 10$~r$_h$) from the center of Cen~I, \textit{Gaia} DR2 \citep{GaiaDR2, Holl2018} flags only one star as ``VARIABLE'' but no pulsation parameters nor proper motions are given for this star. When matching with our catalog, this star turned out to be V4, a $\delta$~Sct/SX~Phe from the MW field. The mean value of $G$ for this star is 20.14 mag, which is consistent with the mean $g$ magnitude we obtain.

Using \textit{Gaia}  Early Data Release 3  \citep[EDR3;][]{GaiaEDR3}, we find proper motions for three of our variable stars, the two $\delta$~Sct/SX~Phe stars (V4 and V5), and one of the Cen~I RRL stars (CenI-V2). The remaining two Cen~I RRL are either not in the catalog (CenI-V3) or do not have an astrometric solution (CenI-V1). Table~\ref{tab:pm} lists the \textit{Gaia} EDR3 \verb+source_id+ and proper motions for these stars.

\begin{figure}
    \hspace{-0.5cm}
    \includegraphics[width=1.2\columnwidth]{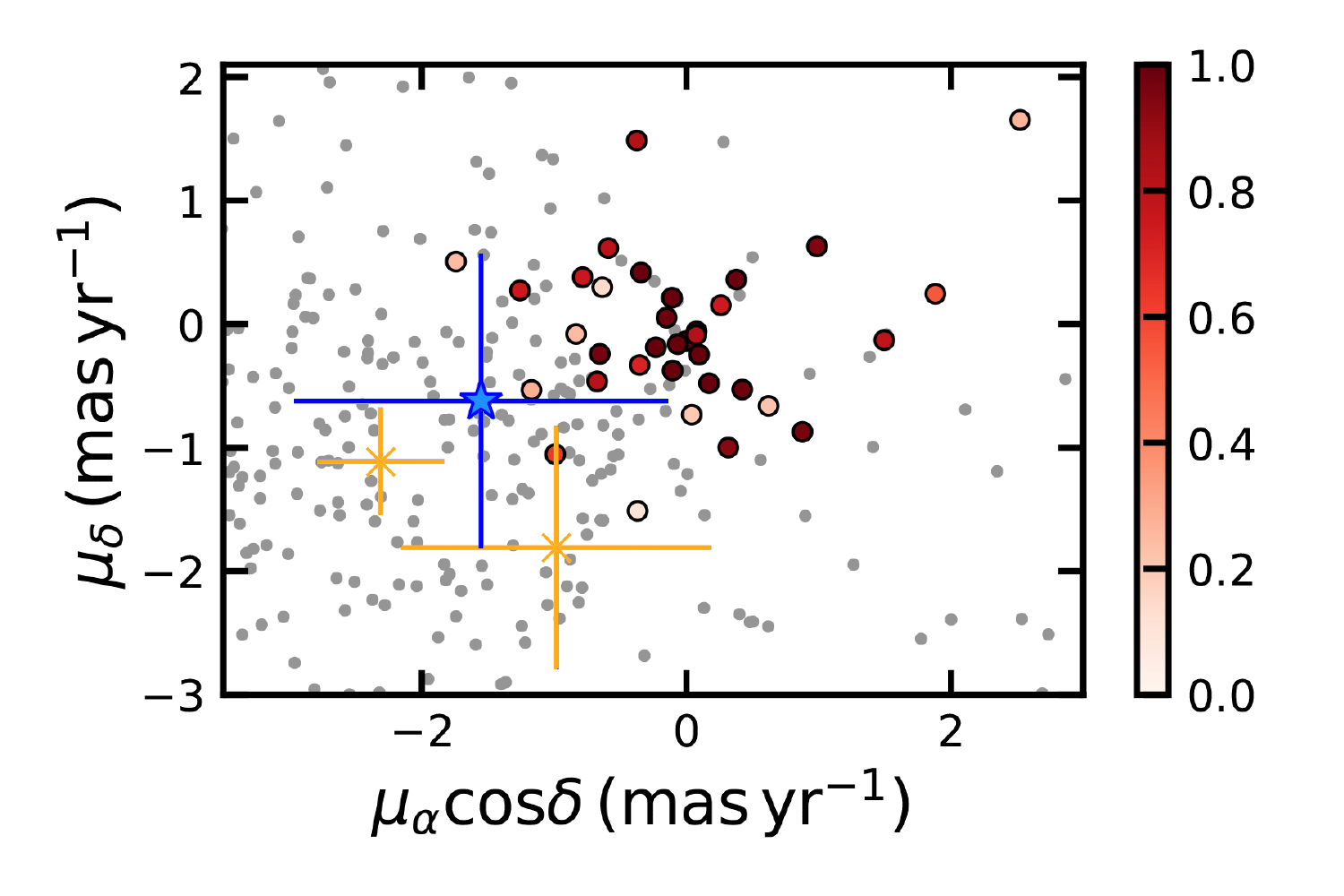}
    \caption{Proper motions from \textit{Gaia} EDR3 of Cen~I field. The grey points represent the proper motions of the field stars nearby (within $30\arcmin$) and consistent with an old, metal-poor isochrone (see \S~\ref{sec:gaia} for more details). The reddish dots represent the candidate RGB members. The
    membership probability is shown in the colorbar. The blue star is the proper motion of the RRL star CenI-V2 and the orange crosses denote the proper motions of the two MW field $\delta$~Sct/SX~Phe stars.}
    \label{fig:pm}
\end{figure}

In Figure~\ref{fig:pm}, we compare the proper motion for CenI-V2 and the two $\delta$~Sct/SX~Phe stars with candidate red-giant branch (RGB) members of Cen~I (red) and MW foreground stars (grey points).  
The candidate Cen~I and MW foreground stars here are selected in a similar manner to \citet{Pace2019} and \citet{Mau2020} but updated with {\it Gaia} EDR3 astrometry. Briefly, stars are selected based on their location in the CMD, zero parallax, and small proper motions. 
The remaining stars are used as the input to a proper motion and spatial mixture model to identify the Cen~I proper motion and candidate members.  More details can be found in \citet{Pace2019} and Pace et al. (in prep).
The proper motion of CenI-V2 is consistent with the proper motion of Cen~I \citep[][Pace et al. in prep]{McConnachie2020}.
Both $\delta$~Sct/SX~Phe are consistent with the MW foreground and the brighter $\delta$~Sct/SX~Phe, V5, is excluded from being a member of Cen~I at high significance based on its proper motion. 

\input{tables/proper_motions_edr3}

\section{Distance determination}\label{sec:distance}
RRL stars are considered one of the best standard candles for old stellar systems \citep{Beaton2018} since they follow a well-known period-luminosity-metallicity (PLZ) relation. In particular, it is in the near-infrared bands where the PLZ relations show the smallest scatter \citep[see, e.g.,][]{Caceres&Catelan2008, Marconi2015, Neeley2015}. Therefore, we use the pulsational properties obtained from the $i$ and $z$ light curves of the RRL stars discovered in this work to derive precise distances.

We employed the PLZ in $i$ and $z$ given by \citet{Caceres&Catelan2008} to measure the distance moduli to our recently detected RRL stars. The standard uncertainties of these relations are 0.045 mag and 0.037 mag, respectively. For the metallicity, we used the mean metallicity [Fe/H]$= -2.57\pm 0.12$ from Cen~I RGB stars obtained from preliminary results of unpublished spectroscopic measurements (J. D. Simon, private communication). For the $\alpha$ abundance, we used [$\alpha$/Fe]$=0.3\pm0.2$ based on the average values obtained for other UFD galaxies (e.g., \citealt{Pritzl2005:abundance, Ji2019, Simon2019} and references therein). Therefore, considering the previous values and following the relationship between Z, [Fe/H] and [$\alpha$/Fe] from \citet{Salaris2005} (using Z$_\sun$=0.0014, \citealt{Asplund2021}), we obtain Z=0.0001 for Cen~I. It is important to note that the \citet{Caceres&Catelan2008} PLZ relations were obtained in SDSS passbands, therefore a transformation from SDSS to DES was needed. To do so, we used the following transformation equations that were generated in the same way as the transformation equations obtained by the DES Collaboration using matched stars from DES DR2 and SDSS DR13 in Stripe 82 \citep[][Appendix A]{DES_DR2}.

\begin{equation}
i_{SDSS} = i_{DES} -0.029 + 0.361 (i_{DES}-z_{DES})
\end{equation}
\begin{equation}
z_{SDSS} = z_{DES} -0.026 + 0.125 (i_{DES}-z_{DES})
\end{equation}

\noindent The rms for these relations are 0.016 mag and 0.017 mag, respectively. Also, in order to obtain the true distance modulus ($\mu_0$), we corrected the $i$ and $z$-band photometry for dust extinction (see Section~\ref{sec:search}). The absorption coefficients for the RRL stars of Cen~I used in this work can be found in the fist two columns of Table~\ref{tab:distances}.

The distance moduli obtained for the RRL stars in Cen~I are listed in Table~\ref{tab:distances}. The uncertainties of the individual distance moduli were obtained by propagation of errors considering: i) the photometric uncertainty of the mean magnitude (0.02 mag), ii) the dispersion of the filter transformation equation from DES to SDSS (0.02 mag), iii) the dispersion of the PLZ ($\sim$0.04 mag), iv) the reddening uncertainty (which is usually assumed to be 10\%), and v) the uncertainties of 0.1 dex in [Fe/H] and 0.2 dex in [$\alpha$/Fe]. 

\input{tables/distances}

For the three RRL stars that are located right on the zero-age HB (ZAHB), which is very well defined because of the high number of BHB in this UFD galaxy, we decided not to include the dispersion in magnitude due to evolution since in these cases it seems negligible. 

Finally, from the two more confident RRL stars (i.e., CenI-V1 and CenI-V2)\footnote{The two RRc stars within 2~r$_h$ and complete phase coverage in their light curves.}, the distance modulus of Cen~I $\mu_0 = 20.354 \pm 0.002$~mag ($\sigma=0.03$~mag), which translates in a heliocentric distance of D$_\odot = 117.7 \pm 0.1$~kpc ($\sigma=1.6$~kpc), with an associated systematic error of $0.07$~mag ($4$~kpc). This value was assessed by fitting simultaneously the RRL stars and comparing the zero-points obtained from the theoretical and semi-empirical PLZ relationships in $i$ and $z$, following the same approach described in \citet{MartinezVazquez2015}. The inclusion of CenI-V3 in this analysis would only change the final distance modulus by $-0.02$~mag.

With the precise distance presented in this work and the Gaia EDR3 proper motions, the addition of spectroscopic radial velocities would complete the 6D phase space information which can be used to derive an orbit for Cen~I.

\section{The Oosterhoff classification of Cen~I}\label{sec:oosterhoff}

\begin{figure}
	\centering
	\includegraphics[width=1.1\columnwidth]{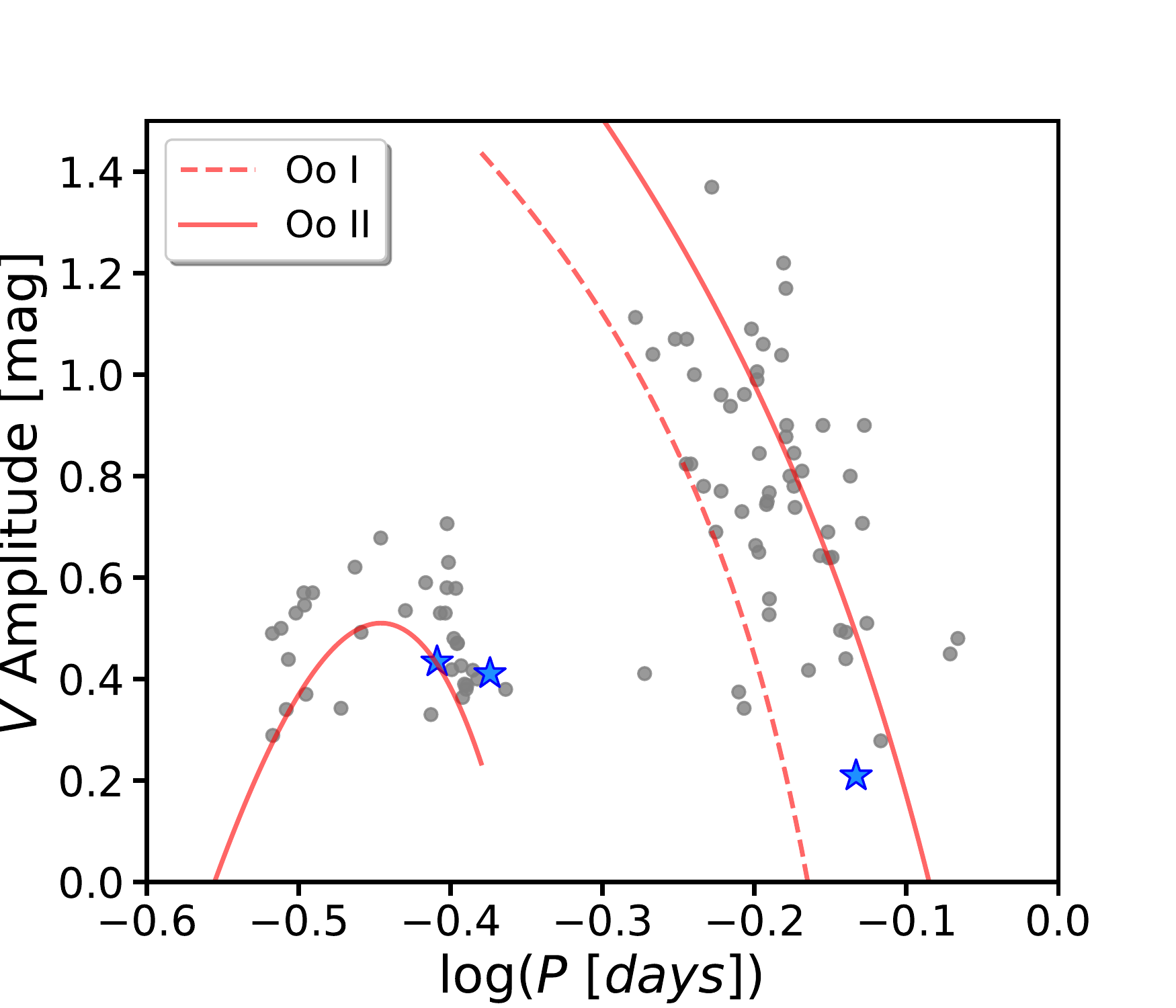}
	\caption{Bailey diagram for the RRL stars found in UFD galaxies with M$_V \gtrsim -6.0$~mag. The RRL stars of Cen~I are highlighted with blue stars. The dashed and solid lines are the locus for the RRL stars in Oo I and Oo II GGCs, respectively.}
	\label{fig:bailey}
\end{figure}

Figure~\ref{fig:bailey} shows period versus amplitude (classically known as the Bailey diagram) for the known RRL stars in UFDs with M$_V \gtrsim -6.0$~mag (see \citealt{MartinezVazquez2019} and \citealt{Vivas2020b} and references therein). The RRL stars of Cen~I are highlighted by blue star symbols. 

It is well known that there is a dichotomy between Galactic globular clusters (GGCs) when studying their mean period and their mean metallicity together, the so-called \textit{Oosterhoff dichotomy} \citep{Oosterhoff1939}. There are two types of GGCs \citep[][and references therein]{Smith1995,Catelan2009, Catelan2015}: \textit{Oosterhoff I} (Oo I) with mean periods for the RRab of $\approx 0.55$~d ($\approx 0.32$~d for the RRc) and mean metallicities between $ -1.3 > \rm{[Fe/H]} > -1.7$, and \textit{Oosterhoff II} (Oo II) with mean periods for the RRab of $\approx 0.65$~d ($\approx 0.37$~d for the RRc) and more metal-poor ($ \rm{[Fe/H]} < -2.0$) systems. 

Figure~\ref{fig:bailey} shows the loci (red curves) provided by \citet{Fabrizio2019} for the RRab stars in Oo I and Oo II type GGCs and that derived by \citet{Kunder2013} for the RRc stars in the cluster M~22, a Oo II GGC. This figure shows how the bulk of RRab stars in UFDs are located around the Oo II line, confirming that UFDs are mainly Oo II systems. Cen I RRL stars seem to overlay well in the Bailey diagram defined by all the UFD RRL stars. While the two Cen~I RRc stars are located near to the Oo II line, the Cen~I RRab star is located between Oo I and Oo II lines (i.e., in the Oosterhoff intermediate region). Therefore, due to such small statistic, it is difficult to make any strong statement about Oosterhoof classification of this galaxy.

Within the context of merger scenarios for the assembly of the Milky Way \citep{Searle1978}, the halo formed from the disruption of small galaxies. The properties of the RRL stars in the halo, which are predominantly Oo I \citep[see e.g., Figure 5 in][]{Drake2013}, does not match those found in the majority of satellites, except for a few of the more massive and metal rich systems \citep[][]{Zinn2014, Fiorentino2015a, Fiorentino2017}. In particular, the fainter dwarf systems (which contain only a few RRL stars) seem to belong preferentially to the Oo II group, therefore it is clear that UFD galaxies are far from being the main contributors to the Galactic halo \citep[e.g., ][]{Vivas2016a, Vivas2020b}.

\section{The frequency of first overtone RRL stars in UFD galaxies}\label{sec:rrc}

As can be seen in Figure~\ref{fig:cmd}, Cen~I hosts a noticeable population of BHB stars. In addition, two of the three RRL stars in Cen~I are RRc stars. Theoretical models predict that RRc stars are preferentially located in the blue edge of the instability strip in the HB \citep[e.g.,][]{Bono1995}. Since most of the UFDs have a noticeable BHB population, we wanted to investigate in this section if UFDs present higher ratio of RRc than more massive dwarf galaxies.

In order to check whether there is a higher frequency of RRc stars in UFD galaxies, we study this ratio individually in all the MW galaxies that have RRL studies so far (see Table 6 in \citealt{MartinezVazquez2017} for the classical dwarf galaxies and Table A1 in \citealt{MartinezVazquez2019} and updates in \citealt{Vivas2020b} for the UFDs). We note that in this analysis we include RRc and RRd stars due to the fact that some studies are not able to distinguish between them. We refer to them as RRcd stars hereafter. 

\begin{figure}
	\hspace{-0.7cm}
	\includegraphics[width=1.1\columnwidth]{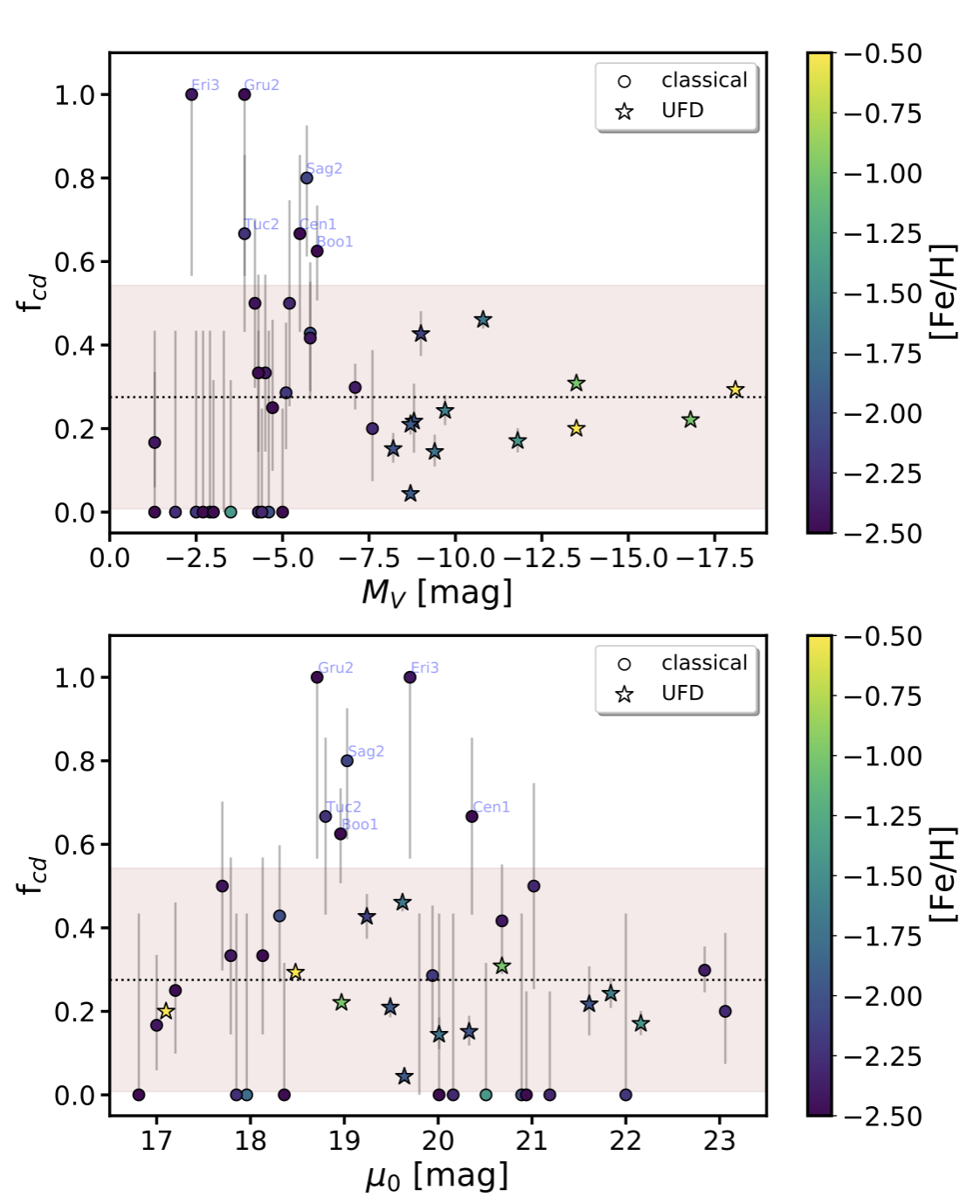}
	\caption{\textit{Top.} Frequency of RRcd stars ($f_{cd}$) in MW satellite galaxies versus their absolute magnitude (M$_V$), color coded by the mean metallicity of the host galaxy. \textit{Bottom.} Same as top panel but as a function of the true distance modulus. The dotted line shows the average of the $f_{cd}$ values. The shaded area represents the 3$\sigma$ region and the grey error bars are the Bayesian errors associated to the $f_{cd}$ values. Classic and UFD galaxies are represented by different symbols as shown in the legend.} 
	\label{fig:ratio}
\end{figure}

Figure~\ref{fig:ratio} shows the frequency of RRcd stars, $f_{cd} = N_{RRcd}/(N_{RRab}+N_{RRcd})$, found in MW dwarf galaxies versus their absolute magnitude (top panel) and their distance moduli (bottom panel). The error bars are the Bayesian errors associated to the $f_{cd}$ values, obtained following \citet{Paterno2004}. We color coded the data based on the mean metallicity of the dwarf galaxy. It is clear that there is no particular trend associated with the mass, distance or metallicity of the host galaxy. The average value of $f_{cd}$ is 0.28 (dashed line), with a dispersion of 0.27 (shaded region). We can see that there are several outliers with $f_{cd} \ga 0.50$: Bootes~I, Cen~I, Tucana~II, Sagittarius~II, Grus~II, and Eridanus~III. All of them are UFD galaxies and have metallicities of ${\rm [Fe/H]} < -2.1$ dex. On the other hand we see that there are 12 UFD galaxies that do not contain any RRcd stars. However, we can see in Figure~\ref{fig:ratio} that there is no indication that UFD galaxies have a higher ratio of RRcd stars than classical dwarf galaxies and that the outliers ($f_{cd} \ga 0.50$) occur only for UFDs, not for classical dwarfs. In addition, the frequency of RRcd and RRab stars in the UFD galaxies is strongly dominated by the small number of RRL stars that belong to them. This is reflected in the dispersion of the $f_{cd}$, 0.31 for the UFDs while for the classical dwarf galaxies is only 0.11. Most of the UFDs that have either a high or null frequency of RRcd stars harbor fewer than 5 RRL stars. Outliers in $f_{cd}$ are also observed among GGCs \citep{Fabrizio2021}.

On the other hand, if we combine all the MW UFDs, the mean $f_{cd}$ is 0.29, which is similar to the one obtained for the classical MW dwarfs ($f_{cd}$ = 0.24). This indicates that frequency of RRcd stars is consistent between UFD and classical dwarf galaxies. 

\section{On the extension of Cen~I}\label{sec:extension}

\begin{figure}
    \hspace{-0.7cm}
    \includegraphics[width=1.05\columnwidth]{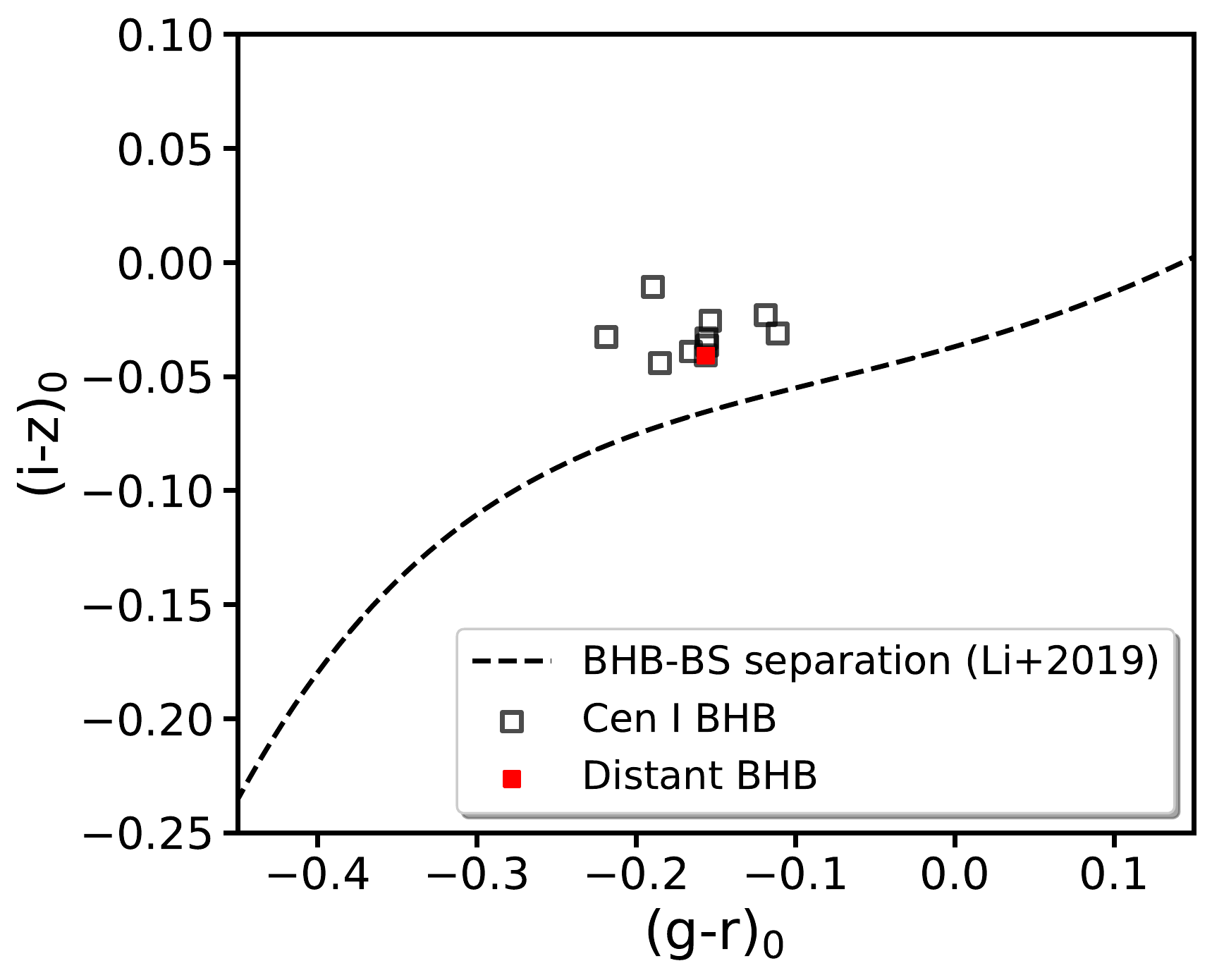}
    \caption{Color-color distribution of BHB stars in Cen~I. The dashed curve shows the polynomial in $(g-r)_0$ versus $(i-z)_0$ proposed by \citet{Li2019} to separate BHB from BS stars.}
    \label{fig:color2}
\end{figure}

Two of the three discovered RRL stars in Cen~I are located within 2~r$_h$, at $2.4\arcmin$ and $3.4\arcmin$, while the third RRL star (CenI-V3) is at $13.1\arcmin$ ($\sim6$\,r$_h$) (see Figure~\ref{fig:spatial}). In addition, out of the ten BHB candidates in Cen~I, nine are centrally concentrated in the inner 3\,r$_h$ (see black squares in Figure~\ref{fig:spatial}) but the remaining one is located much farther out ($21.4\arcmin$). In order to check whether these stars are BHB stars at the distance of Cen~I or foreground blue straggler (BS) stars, we use our $giz$ photometry plus the $r$ photometry from DELVE DR1 and check their positions in the $(g-r)_0$ versus $(i-z)_0$ plane (see Figure~\ref{fig:color2}). Thanks to the BHB-BS separation obtained by \citet[][their equation 5]{Li2019}, we see that indeed all of them are in the region of the color-color space where BHB stars are supposed to be (even the BHB with the largest angular separation).

Both the distant BHB star (located at $21.4\arcmin$) and the CenI-V3 RRL star (at $13.1\arcmin$) are located along the major axis of Cen~I but in opposite directions, which could be possible evidence of a tidal disruption event (see Figure~\ref{fig:spatial}). In the absence of perturbations (e.g. from the LMC) the disruption direction should align with the proper motion vector on the sky. The reflex-corrected proper motion of Cen~I (arrow in Figure~\ref{fig:spatial}) using the $Gaia$ EDR3 proper motion of Cen~I from \citet{McConnachie2020}, the distance from this paper, and positions from \citet{Mau2020} is ($\mu_{\alpha}\cos{\delta}$, $\mu_{\delta}$) = ($+0.11$, $-0.06$) mas yr$^{-1}$. This is roughly perpendicular to the position angle and would naively argue against tidal disruption as an explanation for the positions of CenI-V3 and the distant BHB star.

For more insight, we estimated the tidal radius of Cen~I. From preliminary analysis of unpublished spectroscopic measurements (J. D. Simon, private communication), the velocity dispersion of Cen~I is 5.5 km~s$^{-1}$. For this velocity dispersion, the mass within the half-light radius (using the formula from \citealt{Wolf2010}) is $2.2\times10^6$ M$_\sun$. Using that mass and the MW potential from \citet{CarlinSand2018} in the equation for the Jacobi radius from \citet{BinneyTremaine2008}, the tidal radius of Cen~I is 1 kpc (i.e., $29\arcmin$). Since this assumed mass for Cen~I is very conservative relying only on the dynamical mass in the central regions of the galaxy, this estimate of the tidal radius can be regarded as a lower limit. Therefore, we conclude that the most distant Cen~I RRL star and the most distant BHB star are inside the tidal radius of Cen~I. Deeper imaging reaching several magnitudes below the main-sequence turn off and spectroscopic studies of Cen~I will be required to perform a more detailed characterization of its outer regions and possible tidal extension.

\section{Conclusions} \label{sec:conclusions}

We present in this work the first study of variable stars in the recently discovered UFD galaxy Cen~I. From multi-epoch $giz$ DECam observations, we discover three RRL stars in Cen~I and we detect two $\delta$~Sct/SX~Phe belonging to the MW field.

Two of them are first overtone (RRc) stars and the remaining one is a fundamental pulsator (RRab) star. The two RRc stars are located within 2\,r$_h$ while the RRab star (CenI-V3) is at $\sim6$\,r$_h$. From a smooth distribution of Galactic halo RRL stars, it is not expected to find three MW halo RRL stars clumped together in space at these large distances. In particular, from the density profile of RRL stars derived in \citet{Medina2018}, the significance of having three or more MW halo RRL that could contaminate Cen~I HB is 4.7$\sigma$. Therefore, we conclude that these three RRL stars found are high confidence members of Cen~I. 

We measure a distance modulus for Cen~I of $\mu_0 = 20.354 \pm 0.002$~mag ($\sigma=0.03$~mag), a heliocentric distance of D$_\odot = 117.7 \pm 0.1$~kpc ($\sigma=1.6$~kpc), based on its best sampled RRL stars (i.e., the two RRc stars). The systematic error associated to this measurement due to the uncertainties on the photometry, reddening, [Fe/H] and [$\alpha$/Fe], is of $0.07$~mag ($4$~kpc). This distance measurement agrees with the distance obtained in the discovery paper by \citet[][$20.3 \pm 0.1$~mag]{Mau2020}.

The frequency of RRcd stars in MW dwarf galaxies has a mean value of 0.28 with no trend with the M$_V$, $\mu_0$, or [Fe/H]. Some UFDs, including Cen~I, present higher RRcd ratios ($f_{cd} \ga 0.5$), although no strong conclusions can be drawn for individual UFDs due to limited statistics. However, if we combine all the UFDs, the ratio of RRcd is similar to the one obtained for the classical dwarfs ($f_{cd}$ $\sim$ 0.3). Therefore, the fraction of RRcd stars is consistent between UFD and classical dwarf galaxies.

The location of the Cen~I RRL stars in the Bailey diagram is in good agreement with general location of RRL stars from UFD galaxies. Comparing the properties of the RRL stars in UFDs (mainly Oo II) and those from the halo of the MW (mainly Oo I), it is clear that UFDs are far from being the main contributors to the Galactic halo \citep{Vivas2020b}. Nevertheless, since UFDs are some of the most ancient systems in the Universe, they can also help us to better understand the hierarchical formation and evolution of our Galaxy. 

With the advent of the Vera C. Rubin Legacy Survey of Space and Time (LSST, \citealt{Ivezic2019}), numerous ultra-faint systems will be discovered. The detection of RRL stars and their role as standard candles is crucial to measure accurate distances to UFDs. This, in combination with proper motions and spectroscopic data will allow us to derive their orbits. Thus, time-domain studies of UFDs are necessary to help address questions about their nature, formation, evolution, and contribution to the Galactic halo.

\acknowledgments

We thank our anonymous referee for the comments and suggestions that have helped to improve the content of this paper.
\par The DELVE project is partially supported by Fermilab LDRD project L2019-011 and the NASA Fermi Guest Investigator Program Cycle 9 No. 91201.
\par This work was supported in part by the U.S. Department of Energy, Office of Science, Office of Workforce Development for Teachers and Scientists (WDTS) under the Science Undergraduate Laboratory Internships (SULI) program. 
\par This project used data obtained with the Dark Energy Camera (DECam), which was constructed by the Dark Energy Survey (DES) collaboration. Funding for the DES Projects has been provided by the US Department of Energy, the US National Science Foundation, the Ministry of Science and Education of Spain, the Science and Technology Facilities Council of the United Kingdom, the Higher Education Funding Council for England, the National Center for Supercomputing Applications at the University of Illinois at Urbana-Champaign, the Kavli Institute for Cosmological Physics at the University of Chicago, Center for Cosmology and Astro-Particle Physics at the Ohio State University, the Mitchell Institute for Fundamental Physics and Astronomy at Texas A\&M University, Financiadora de Estudos e Projetos, Funda\c{c}\~{a}o Carlos Chagas Filho de Amparo \`a Pesquisa do Estado do Rio de Janeiro, Conselho Nacional de Desenvolvimento Científico e Tecnológico and the Minist\'erio da Ci\^encia, Tecnologia e Inova\c{c}\~{a}o, the Deutsche Forschungsgemeinschaft and the Collaborating Institutions in the Dark Energy Survey.
\par The Collaborating Institutions are Argonne National Laboratory, the University of California at Santa Cruz, the University of Cambridge, Centro de Investigaciones Energ\'eticas, Medioambientales y Tecnol\'ogicas–Madrid, the University of Chicago, University College London, the DES-Brazil Consortium, the University of Edinburgh, the Eidgen\"ossische Technische Hochschule (ETH) Z\"urich, Fermi National Accelerator Laboratory, the University of Illinois at Urbana-Champaign, the Institut de Ci\`encies de l'Espai (IEEC/CSIC), the Institut de F\'isica d'Altes Energies, Lawrence Berkeley National Laboratory, the Ludwig-Maximilians Universit\"at M\"unchen and the associated Excellence Cluster Universe, the University of Michigan, NSF's NOIRLab, the University of Nottingham, the Ohio State University, the OzDES Membership Consortium, the University of Pennsylvania, the University of Portsmouth, SLAC National Accelerator Laboratory, Stanford University, the University of Sussex, and Texas A\&M University.
\par Based on observations at Cerro Tololo Inter-American Observatory, NSF’s NOIRLab (NOIRLab Prop. ID 2020A-0238; PI: C.~E.~Mart\'inez-V\'azquez), which is managed by the Association of Universities for Research in Astronomy (AURA) under a cooperative agreement with the National Science Foundation.
\par This manuscript has been authored by Fermi Research Alliance, LLC under Contract No. DE-AC02-07CH11359 with the U.S. Department of Energy, Office of Science, Office of High Energy Physics. The United States Government retains and the publisher, by accepting the article for publication, acknowledges that the United States Government retains a non-exclusive, paid-up, irrevocable, world-wide license to publish or reproduce the published form of this manuscript, or allow others to do so, for United States Government purposes.
\par This work has made use of data from the European Space Agency (ESA) mission
{\it Gaia} (\url{https://www.cosmos.esa.int/gaia}), processed by the {\it Gaia}
Data Processing and Analysis Consortium (DPAC,
\url{https://www.cosmos.esa.int/web/gaia/dpac/consortium}). Funding for the DPAC
has been provided by national institutions, in particular the institutions
participating in the {\it Gaia} Multilateral Agreement.

\vspace{5mm}
\facilities{Blanco (DECam), \textit{Gaia}}

\software{astropy \citep{Astropy2013}, matplotlib \citep{Matplotlib},
          SourceExtractor \citep{Bertin:1996}
          }

\bibliographystyle{aasjournal}
\input{main.bbl}



\end{document}

%% file: tables/photometry_variables_CenI_tab_red.tex
\begin{table*}
\begin{scriptsize}
\caption{Photometry of the variable stars found in the field of Cen~I}
\label{tab:photometry}
\hspace{0.7cm}
\begin{tabular}{ccccccccc} 
\toprule
MJD$_g$    &    $g$    &    $\sigma_g$    &    MJD$_i$    &    $i$    &    $\sigma_i$    &    MJD$_z$    &    $z$    &    $\sigma_z$   \\
\midrule
\multicolumn{9}{c}{CenI-V1} \\
\midrule
   56373.2754  &   20.987  &    0.014  &   56357.1475  &   20.933  &    0.021  &   56357.1449  &   20.925  &    0.065 \\
   56373.2768  &   20.984  &    0.015  &   57831.0747  &   20.808  &    0.028  &   56360.1379  &   20.885  &    0.039 \\
   57831.0729  &   21.085  &    0.025  &   57831.0875  &   20.815  &    0.028  &   56360.1389  &   20.852  &    0.038 \\
   57831.0738  &   21.087  &    0.025  &   58888.2045  &   20.800  &    0.027  &   56361.1419  &   20.809  &    0.031 \\
   57831.0858  &   21.045  &    0.025  &   58888.2451  &   20.803  &    0.024  &   56361.1430  &   20.807  &    0.031 \\
   57831.0866  &   21.067  &    0.026  &   58888.2874  &   20.849  &    0.021  &   56373.2729  &   20.784  &    0.040 \\
   58888.2020  &   20.903  &    0.047  &   58888.3292  &   20.972  &    0.024  &   56373.2792  &   20.783  &    0.042 \\
   58888.2426  &   20.992  &    0.039  &   58889.2138  &   21.092  &    0.030  &   57849.0516  &   20.719  &    0.045 \\
   58888.2850  &   21.315  &    0.050  &   58889.2555  &   21.070  &    0.026  &   57850.0564  &   20.990  &    0.035 \\
   58888.3268  &   21.378  &    0.047  &   58889.2961  &   20.908  &    0.022  &   57850.0581  &   21.029  &    0.044 \\
   ... &  ... & ... & ... & ... & ... & ... & ... & ... \\
\bottomrule
\end{tabular}
\tablecomments{MJD is the Modified Julian Date of mid-exposure. This table is a portion of its entirety, which will be available in the online journal.}
\end{scriptsize}
\end{table*}

%% file: tables/rrl_prop.tex
\begin{table*}
\caption{Coordinates, pulsation properties and average photometry of the variable stars in Cen~I.}
\label{tab:rrl_prop}
\hspace{-2cm}
\begin{tabular}{ccccccccrrrc}
\toprule
Star & RA & DEC  & r\tablenotemark{$^{a}$} & Period & $g$ & $i$ & $z$ & $\Delta g$ \tablenotemark{$^{b}$} & $\Delta i$ \tablenotemark{$^{b}$} & $\Delta z$ \tablenotemark{$^{b}$} & Type \\
& (degrees) & (degrees) & (arcmin) & (days) & (mag)  & (mag)  & (mag)  & (mag)  & (mag) & (mag)  &   \\
\midrule
CenI-V1 & 189.570323 & $-$40.939879 &  2.38 & 0.3899181 & 21.24 & 20.92 & 20.86 & 0.56 & 0.25 & 0.24 & RRc  \\
CenI-V2 & 189.633635 & $-$40.878072 &  3.37 & 0.4224812 & 21.20 & 20.88 & 20.83 & 0.53 & 0.27 & 0.20 & RRc \\
CenI-V3 & 189.584351 & $-$41.101214 & 13.12 & 0.7358982 & 21.21 & 20.67 & 20.59 & $>$0.27 & $>$ 0.36 & $>$0.26 & RRab \\
V4      & 189.516808 & $-$40.744898 & 12.86 & 0.0413440 & 20.96 & 20.82 & 20.81 & 0.37 & 0.19 & 0.13 & $\delta$~Sct/SX Phe\tablenotemark{$^{c}$} \\
V5      & 189.776219 & $-$40.977123 & 16.19 & 0.0724028 & 20.31 & 20.12 & 20.11 & 0.72 & 0.37 & 0.33 & $\delta$~Sct/SX Phe\tablenotemark{$^{c}$} \\
\bottomrule
\end{tabular}
\tablenotetext{a}{r is the elliptical radius measured from each star to the center of Cen~I.}

\tablenotetext{b}{$\Delta_{band}$ refers to the amplitude of the variable star in a particular band.}
\tablenotetext{c}{Milky Way field stars.}
\end{table*}

%% file: tables/proper_motions_edr3.tex
\begin{table}
\caption{\textit{Gaia} EDR3 proper motions for the variable stars discovered in the field of Cen~I.}
\label{tab:pm}
\centering
\setlength{\tabcolsep}{2pt}
\begin{tabular}{lccc}
\toprule
  Star    &  \textit{Gaia} source\_id         &  $\mu_{\alpha}\cos{\delta}$   &  $\mu_{\delta}$      \\
          &                      &   (mas yr$^{-1}$)             &   (mas yr$^{-1}$)    \\
\midrule          
  CenI-V1 & 6146232551449525376  &      ---                      &       ---            \\  
  CenI-V2 & 6146234235076699392  &  $-$1.55 $\pm$ 1.41           &  $-$0.62 $\pm$ 1.19  \\  
  CenI-V3 &     ---              &      ---                      &       ---            \\ 
  V4      & 6146250826534361472  &  $-$0.98 $\pm$ 1.17           &  $-$1.80 $\pm$ 0.98  \\  
  V5      & 6146230004532587264  &  $-$2.31 $\pm$ 0.48           &  $-$1.11 $\pm$ 0.43  \\  
\bottomrule
\end{tabular}
\end{table}

%% file: tables/distances.tex
\begin{table*}
\caption{Distance moduli of the RRL stars in Cen~I.}
\label{tab:distances}
\hspace{-1.3cm}
\begin{tabular}{lcccccccc}
\toprule
Star & A$_i$ & A$_z$ & $\mu_0$ (PLZ$_i$) & D$_{\odot}$ (PLZ$_i$) & $\mu_0$ (PLZ$_z$) & D$_{\odot}$ (PLZ$_z$) & $\langle \mu_0 \rangle$ & $\langle$D$_{\odot}\rangle$ \\
& (mag) & (mag)  & (mag) & (kpc)  & (mag) & (kpc)  & (mag) & (kpc) \\
\midrule
Cen~I-V1 & 0.192 & 0.146 & 20.38 $\pm$ 0.07 & 119 $\pm$ 4 & 20.33 $\pm$ 0.06 & 116 $\pm$ 3 & 20.35 $\pm$ 0.07 & 117 $\pm$ 4\\
Cen~I-V2 & 0.203 & 0.155 & 20.37 $\pm$ 0.07 & 119 $\pm$ 4 & 20.34 $\pm$ 0.06 & 117 $\pm$ 3 & 20.35 $\pm$ 0.07 & 117 $\pm$ 4\\
Cen~I-V3 & 0.177 & 0.135 & 20.31 $\pm$ 0.07 & 115 $\pm$ 4 & 20.27 $\pm$ 0.06 & 113 $\pm$ 3 & 20.29 $\pm$ 0.07 & 114 $\pm$ 4\\
\bottomrule
\end{tabular}
\tablecomments{The last two columns are the final distance moduli and heliocentric distances for the RRL stars obtained by averaging the PLZ$_i$ and PLZ$_z$ values for each star.}  
\end{table*}

%% file: main.bbl
 \newcommand{\noop}[1]{}

%% file: main.bbl
\begin{thebibliography}{}
\expandafter\ifx\csname natexlab\endcsname\relax\def\natexlab#1{#1}\fi
\providecommand{\url}[1]{\href{#1}{#1}}
\providecommand{\dodoi}[1]{doi:~\href{http://doi.org/#1}{\nolinkurl{#1}}}
\providecommand{\doeprint}[1]{\href{http://ascl.net/#1}{\nolinkurl{http://ascl.net/#1}}}
\providecommand{\doarXiv}[1]{\href{https://arxiv.org/abs/#1}{\nolinkurl{https://arxiv.org/abs/#1}}}

\bibitem[{{Asplund} {et~al.}(2021){Asplund}, {Amarsi}, \&
  {Grevesse}}]{Asplund2021}
{Asplund}, M., {Amarsi}, A.~M., \& {Grevesse}, N. 2021, arXiv e-prints,
  arXiv:2105.01661.
\newblock \doarXiv{2105.01661}

\bibitem[{{Astropy Collaboration} {et~al.}(2013){Astropy Collaboration},
  {Robitaille}, {Tollerud}, {Greenfield}, {Droettboom}, {Bray}, {Aldcroft},
  {Davis}, {Ginsburg}, {Price-Whelan}, {Kerzendorf}, {Conley}, {Crighton},
  {Barbary}, {Muna}, {Ferguson}, {Grollier}, {Parikh}, {Nair}, {Unther},
  {Deil}, {Woillez}, {Conseil}, {Kramer}, {Turner}, {Singer}, {Fox}, {Weaver},
  {Zabalza}, {Edwards}, {Azalee Bostroem}, {Burke}, {Casey}, {Crawford},
  {Dencheva}, {Ely}, {Jenness}, {Labrie}, {Lim}, {Pierfederici}, {Pontzen},
  {Ptak}, {Refsdal}, {Servillat}, \& {Streicher}}]{Astropy2013}
{Astropy Collaboration}, {Robitaille}, T.~P., {Tollerud}, E.~J., {et~al.} 2013,
  \aap, 558, A33, \dodoi{10.1051/0004-6361/201322068}

\bibitem[{{Beaton} {et~al.}(2018){Beaton}, {Bono}, {Braga}, {Dall'Ora},
  {Fiorentino}, {Jang}, {Mart{\'\i}nez-V{\'a}zquez}, {Matsunaga}, {Monelli},
  {Neeley}, \& {Salaris}}]{Beaton2018}
{Beaton}, R.~L., {Bono}, G., {Braga}, V.~F., {et~al.} 2018, \ssr, 214, 113,
  \dodoi{10.1007/s11214-018-0542-1}

\bibitem[{{Bechtol} {et~al.}(2015){Bechtol}, {Drlica-Wagner}, {Balbinot},
  {Pieres}, {Simon}, {Yanny}, {Santiago}, {Wechsler}, {Frieman}, {Walker},
  {Williams}, {Rozo}, {Rykoff}, {Queiroz}, {Luque}, {Benoit-L{\'e}vy},
  {Tucker}, {Sevilla}, {Gruendl}, {da Costa}, {Fausti Neto}, {Maia}, {Abbott},
  {Allam}, {Armstrong}, {Bauer}, {Bernstein}, {Bernstein}, {Bertin}, {Brooks},
  {Buckley-Geer}, {Burke}, {Carnero Rosell}, {Castander}, {Covarrubias},
  {D'Andrea}, {DePoy}, {Desai}, {Diehl}, {Eifler}, {Estrada}, {Evrard},
  {Fernandez}, {Finley}, {Flaugher}, {Gaztanaga}, {Gerdes}, {Girardi},
  {Gladders}, {Gruen}, {Gutierrez}, {Hao}, {Honscheid}, {Jain}, {James},
  {Kent}, {Kron}, {Kuehn}, {Kuropatkin}, {Lahav}, {Li}, {Lin}, {Makler},
  {March}, {Marshall}, {Martini}, {Merritt}, {Miller}, {Miquel}, {Mohr},
  {Neilsen}, {Nichol}, {Nord}, {Ogando}, {Peoples}, {Petravick}, {Plazas},
  {Romer}, {Roodman}, {Sako}, {Sanchez}, {Scarpine}, {Schubnell}, {Smith},
  {Soares-Santos}, {Sobreira}, {Suchyta}, {Swanson}, {Tarle}, {Thaler},
  {Thomas}, {Wester}, {Zuntz}, \& {DES Collaboration}}]{Bechtol2015}
{Bechtol}, K., {Drlica-Wagner}, A., {Balbinot}, E., {et~al.} 2015, \apj, 807,
  50, \dodoi{10.1088/0004-637X/807/1/50}

\bibitem[{{Bernstein} {et~al.}(2018){Bernstein}, {Abbott}, {Armstrong},
  {Burke}, {Diehl}, {Gruendl}, {Johnson}, {Li}, {Rykoff}, {Walker}, {Wester},
  \& {Yanny}}]{Bernstein:2018}
{Bernstein}, G.~M., {Abbott}, T.~M.~C., {Armstrong}, R., {et~al.} 2018, \pasp,
  130, 054501, \dodoi{10.1088/1538-3873/aaa753}

\bibitem[{{Bertin}(2011)}]{Bertin:2011}
{Bertin}, E. 2011, in Astronomical Society of the Pacific Conference Series,
  Vol. 442, Astronomical Data Analysis Software and Systems XX, ed. I.~N.
  {Evans}, A.~{Accomazzi}, D.~J. {Mink}, \& A.~H. {Rots}, 435

\bibitem[{{Bertin} \& {Arnouts}(1996)}]{Bertin:1996}
{Bertin}, E., \& {Arnouts}, S. 1996, \aaps, 117, 393

\bibitem[{{Binney} \& {Tremaine}(2008)}]{BinneyTremaine2008}
{Binney}, J., \& {Tremaine}, S. 2008, {Galactic Dynamics: Second Edition}

\bibitem[{{Bono} {et~al.}(1995){Bono}, {Caputo}, \& {Marconi}}]{Bono1995}
{Bono}, G., {Caputo}, F., \& {Marconi}, M. 1995, \aj, 110, 2365,
  \dodoi{10.1086/117694}

\bibitem[{{Bose} {et~al.}(2018){Bose}, {Deason}, \& {Frenk}}]{Bose2018}
{Bose}, S., {Deason}, A.~J., \& {Frenk}, C.~S. 2018, \apj, 863, 123,
  \dodoi{10.3847/1538-4357/aacbc4}

\bibitem[{{C{\'a}ceres} \& {Catelan}(2008)}]{Caceres&Catelan2008}
{C{\'a}ceres}, C., \& {Catelan}, M. 2008, \apjs, 179, 242,
  \dodoi{10.1086/591231}

\bibitem[{{Carlin} \& {Sand}(2018)}]{CarlinSand2018}
{Carlin}, J.~L., \& {Sand}, D.~J. 2018, \apj, 865, 7,
  \dodoi{10.3847/1538-4357/aad8c1}

\bibitem[{{Catelan}(2009)}]{Catelan2009}
{Catelan}, M. 2009, \apss, 320, 261, \dodoi{10.1007/s10509-009-9987-8}

\bibitem[{{Catelan} \& {Smith}(2015)}]{Catelan2015}
{Catelan}, M., \& {Smith}, H.~A. 2015, {Pulsating Stars}

\bibitem[{{Chambers} {et~al.}(2016){Chambers}, {Magnier}, {Metcalfe},
  {Flewelling}, {Huber}, {Waters}, {Denneau}, {Draper}, {Farrow}, {Finkbeiner},
  {Holmberg}, {Koppenhoefer}, {Price}, {Saglia}, {Schlafly}, {Smartt},
  {Sweeney}, {Wainscoat}, {Burgett}, {Grav}, {Heasley}, {Hodapp}, {Jedicke},
  {Kaiser}, {Kudritzki}, {Luppino}, {Lupton}, {Monet}, {Morgan}, {Onaka},
  {Stubbs}, {Tonry}, {Banados}, {Bell}, {Bender}, {Bernard}, {Botticella},
  {Casertano}, {Chastel}, {Chen}, {Chen}, {Cole}, {Deacon}, {Frenk},
  {Fitzsimmons}, {Gezari}, {Goessl}, {Goggia}, {Goldman}, {Grebel}, {Hambly},
  {Hasinger}, {Heavens}, {Heckman}, {Henderson}, {Henning}, {Holman}, {Hopp},
  {Ip}, {Isani}, {Keyes}, {Koekemoer}, {Kotak}, {Long}, {Lucey}, {Liu},
  {Martin}, {McLean}, {Morganson}, {Murphy}, {Nieto-Santisteban}, {Norberg},
  {Peacock}, {Pier}, {Postman}, {Primak}, {Rae}, {Rest}, {Riess}, {Riffeser},
  {Rix}, {Roser}, {Schilbach}, {Schultz}, {Scolnic}, {Szalay}, {Seitz},
  {Shiao}, {Small}, {Smith}, {Soderblom}, {Taylor}, {Thakar}, {Thiel},
  {Thilker}, {Urata}, {Valenti}, {Walter}, {Watters}, {Werner}, {White},
  {Wood-Vasey}, \& {Wyse}}]{Chambers:2016}
{Chambers}, K.~C., {Magnier}, E.~A., {Metcalfe}, N., {et~al.} 2016, ArXiv
  e-prints.
\newblock \doarXiv{1612.05560}

\bibitem[{{DES Collaboration} {et~al.}(2018){DES Collaboration}, {Abbott},
  {Abdalla}, {Allam}, {et~al.}}]{DR1:2018}
{DES Collaboration}, {Abbott}, T.~M.~C., {Abdalla}, F.~B., {Allam}, S.,
  {et~al.} 2018, \apjs, 239, 18, \dodoi{10.3847/1538-4365/aae9f0}

\bibitem[{{DES Collaboration} {et~al.}(2021){DES Collaboration}, {Abbott},
  {Adamow}, {Aguena}, {Allam}, {Amon}, {Annis}, {Avila}, {Bacon}, {Banerji},
  {Bechtol}, {Becker}, {Bernstein}, {Bertin}, {Bhargava}, {Bridle}, {Brooks},
  {Burke}, {Carnero Rosell}, {Carrasco Kind}, {Carretero}, {Castander},
  {Cawthon}, {Chang}, {Choi}, {Conselice}, {Costanzi}, {Crocce}, {da Costa},
  {Davis}, {De Vicente}, {DeRose}, {Desai}, {Diehl}, {Dietrich},
  {Drlica-Wagner}, {Eckert}, {Elvin-Poole}, {Everett}, {Evrard}, {Ferrero},
  {Fert{\'e}}, {Flaugher}, {Fosalba}, {Friedel}, {Frieman},
  {Garc{\'\i}a-Bellido}, {Gelman}, {Gerdes}, {Giannantonio}, {Gill}, {Gruen},
  {Gruendl}, {Gschwend}, {Gutierrez}, {Hartley}, {Hinton}, {Hollowood},
  {Honscheid}, {Huterer}, {James}, {Jeltema}, {Johnson}, {Kent}, {Kron},
  {Kuehn}, {Kuropatkin}, {Lahav}, {Li}, {Lidman}, {Lin}, {MacCrann}, {Maia},
  {Manning}, {March}, {Marshall}, {Martini}, {Melchior}, {Menanteau}, {Miquel},
  {Morgan}, {Myles}, {Neilsen}, {Ogando}, {Palmese}, {Paz-Chinch{\'o}n},
  {Petravick}, {Pieres}, {Plazas}, {Pond}, {Rodriguez-Monroy}, {Romer},
  {Roodman}, {Rykoff}, {Sako}, {Sanchez}, {Santiago}, {Serrano},
  {Sevilla-Noarbe}, {Allyn. Smith}, {Smith}, {Soares-Santos}, {Suchyta},
  {Swanson}, {Tarle}, {Thomas}, {To}, {Tremblay}, {Troxel}, {Tucker}, {Turner},
  {Varga}, {Walker}, {Wechsler}, {Weller}, {Wester}, {Wilkinson}, {Yanny},
  {Zhang}, {Nikutta}, {Fitzpatrick}, {Jacques}, {Scott}, {Olsen}, {Huang},
  {Herrera}, {Juneau}, {Nidever}, {Weaver}, {Adean}, {Correia}, {de Freitas},
  {Freitas}, {Singulani}, \& {Vila-Verde}}]{DES_DR2}
{DES Collaboration}, {Abbott}, T.~M.~C., {Adamow}, M., {et~al.} 2021, arXiv
  e-prints, arXiv:2101.05765.
\newblock \doarXiv{2101.05765}

\bibitem[{{Drake} {et~al.}(2013){Drake}, {Catelan}, {Djorgovski}, {Torrealba},
  {Graham}, {Belokurov}, {Koposov}, {Mahabal}, {Prieto}, {Donalek}, {Williams},
  {Larson}, {Christensen}, \& {Beshore}}]{Drake2013}
{Drake}, A.~J., {Catelan}, M., {Djorgovski}, S.~G., {et~al.} 2013, \apj, 763,
  32, \dodoi{10.1088/0004-637X/763/1/32}

\bibitem[{{Drlica-Wagner} {et~al.}(2015){Drlica-Wagner}, {Bechtol}, {Rykoff},
  {Luque}, {Queiroz}, {Mao}, {Wechsler}, {Simon}, {Santiago}, {Yanny},
  {Balbinot}, {Dodelson}, {Fausti Neto}, {James}, {Li}, {Maia}, {Marshall},
  {Pieres}, {Stringer}, {Walker}, {Abbott}, {Abdalla}, {Allam},
  {Benoit-L{\'e}vy}, {Bernstein}, {Bertin}, {Brooks}, {Buckley-Geer}, {Burke},
  {Carnero Rosell}, {Carrasco Kind}, {Carretero}, {Crocce}, {da Costa},
  {Desai}, {Diehl}, {Dietrich}, {Doel}, {Eifler}, {Evrard}, {Finley},
  {Flaugher}, {Fosalba}, {Frieman}, {Gaztanaga}, {Gerdes}, {Gruen}, {Gruendl},
  {Gutierrez}, {Honscheid}, {Kuehn}, {Kuropatkin}, {Lahav}, {Martini},
  {Miquel}, {Nord}, {Ogando}, {Plazas}, {Reil}, {Roodman}, {Sako}, {Sanchez},
  {Scarpine}, {Schubnell}, {Sevilla-Noarbe}, {Smith}, {Soares-Santos},
  {Sobreira}, {Suchyta}, {Swanson}, {Tarle}, {Tucker}, {Vikram}, {Wester},
  {Zhang}, {Zuntz}, \& {DES Collaboration}}]{DrlicaWagner2015}
{Drlica-Wagner}, A., {Bechtol}, K., {Rykoff}, E.~S., {et~al.} 2015, \apj, 813,
  109, \dodoi{10.1088/0004-637X/813/2/109}

\bibitem[{{Drlica-Wagner} {et~al.}(2021){Drlica-Wagner}, {Carlin}, {Nidever},
  {Ferguson}, {Kuropatkin}, {Adam{\'o}w}, {Cerny}, {Choi}, {Esteves},
  {Mart{\'\i}nez-V{\'a}zquez}, {Mau}, {Miller}, {Mutlu-Pakdil}, {Neilsen},
  {Olsen}, {Pace}, {Riley}, {Sakowska}, {Sand}, {Santana-Silva}, {Tollerud},
  {Tucker}, {Vivas}, {Zaborowski}, {Zenteno}, {Abbott}, {Allam}, {Bechtol},
  {Bell}, {Bell}, {Bilaji}, {Bom}, {Carballo-Bello}, {Cioni}, {Diaz-Ocampo},
  {de Boer}, {Erkal}, {Gruendl}, {Hernandez-Lang}, {Hughes}, {James},
  {Johnson}, {Li}, {Mao}, {Mart{\'\i}nez-Delgado}, {Massana}, {McNanna},
  {Morgan}, {Nadler}, {No{\"e}l}, {Palmese}, {Peter}, {Rykoff}, {S{\'a}nchez},
  {Shipp}, {Simon}, {Smercina}, {Soares-Santos}, {Stringfellow}, {Tavangar},
  {van der Marel}, {Walker}, {Wechsler}, {Wu}, {Yanny}, {Fitzpatrick}, {Huang},
  {Jacques}, {Nikutta}, \& {Scott}}]{Drlica-Wagner2021}
{Drlica-Wagner}, A., {Carlin}, J.~L., {Nidever}, D.~L., {et~al.} 2021, arXiv
  e-prints, arXiv:2103.07476.
\newblock \doarXiv{2103.07476}

\bibitem[{{Fabrizio} {et~al.}(2019){Fabrizio}, {Bono}, {Braga}, {Magurno},
  {Marinoni}, {Marrese}, {Ferraro}, {Fiorentino}, {Giuffrida}, {Iannicola},
  {Monelli}, {Altavilla}, {Chaboyer}, {Dall'Ora}, {Gilligan}, {Layden},
  {Marengo}, {Nonino}, {Preston}, {Sesar}, {Sneden}, {Valenti}, {Th{\'e}venin},
  \& {Zoccali}}]{Fabrizio2019}
{Fabrizio}, M., {Bono}, G., {Braga}, V.~F., {et~al.} 2019, \apj, 882, 169,
  \dodoi{10.3847/1538-4357/ab3977}

\bibitem[{{Fabrizio} {et~al.}(2021){Fabrizio}, {Braga}, {Crestani}, {Bono},
  {Ferraro}, {Fiorentino}, {Iannicola}, {Preston}, {Sneden}, {Th{\'e}venin},
  {Altavilla}, {Chaboyer}, {Dall'Ora}, {da Silva}, {Grebel}, {Gilligan},
  {Lala}, {Lemasle}, {Magurno}, {Marengo}, {Marinoni}, {Marrese},
  {Mart{\`\i}nez-V{\`a}zquez}, {Matsunaga}, {Monelli}, {Mullen}, {Neeley},
  {Nonino}, {Prudil}, {Salaris}, {Stetson}, {Valenti}, \&
  {Zoccali}}]{Fabrizio2021}
{Fabrizio}, M., {Braga}, V.~F., {Crestani}, J., {et~al.} 2021, arXiv e-prints,
  arXiv:2107.00919.
\newblock \doarXiv{2107.00919}

\bibitem[{{Fiorentino} {et~al.}(2015){Fiorentino}, {Bono}, {Monelli},
  {Stetson}, {Tolstoy}, {Gallart}, {Salaris}, {Mart{\'{\i}}nez-V{\'a}zquez}, \&
  {Bernard}}]{Fiorentino2015a}
{Fiorentino}, G., {Bono}, G., {Monelli}, M., {et~al.} 2015, \apjl, 798, L12,
  \dodoi{10.1088/2041-8205/798/1/L12}

\bibitem[{{Fiorentino} {et~al.}(2017){Fiorentino}, {Monelli}, {Stetson},
  {Bono}, {Gallart}, {Mart{\'{\i}}nez-V{\'a}zquez}, {Bernard}, {Massari},
  {Braga}, \& {Dall'Ora}}]{Fiorentino2017}
{Fiorentino}, G., {Monelli}, M., {Stetson}, P.~B., {et~al.} 2017, \aap, 599,
  A125, \dodoi{10.1051/0004-6361/201629501}

\bibitem[{{Flaugher} {et~al.}(2015){Flaugher}, {Diehl}, {Honscheid}, {Abbott},
  {Alvarez}, {Angstadt}, {Annis}, {Antonik}, {Ballester}, {Beaufore},
  {Bernstein}, {Bernstein}, {Bigelow}, {Bonati}, {Boprie}, {Brooks},
  {Buckley-Geer}, {Campa}, {Cardiel-Sas}, {Castander}, {Castilla}, {Cease},
  {Cela-Ruiz}, {Chappa}, {Chi}, {Cooper}, {da Costa}, {Dede}, {Derylo},
  {DePoy}, {de Vicente}, {Doel}, {Drlica-Wagner}, {Eiting}, {Elliott}, {Emes},
  {Estrada}, {Fausti Neto}, {Finley}, {Flores}, {Frieman}, {Gerdes},
  {Gladders}, {Gregory}, {Gutierrez}, {Hao}, {Holland}, {Holm}, {Huffman},
  {Jackson}, {James}, {Jonas}, {Karcher}, {Karliner}, {Kent}, {Kessler},
  {Kozlovsky}, {Kron}, {Kubik}, {Kuehn}, {Kuhlmann}, {Kuk}, {Lahav}, {Lathrop},
  {Lee}, {Levi}, {Lewis}, {Li}, {Mandrichenko}, {Marshall}, {Martinez},
  {Merritt}, {Miquel}, {Mu{\~n}oz}, {Neilsen}, {Nichol}, {Nord}, {Ogando},
  {Olsen}, {Palaio}, {Patton}, {Peoples}, {Plazas}, {Rauch}, {Reil}, {Rheault},
  {Roe}, {Rogers}, {Roodman}, {Sanchez}, {Scarpine}, {Schindler}, {Schmidt},
  {Schmitt}, {Schubnell}, {Schultz}, {Schurter}, {Scott}, {Serrano}, {Shaw},
  {Smith}, {Soares-Santos}, {Stefanik}, {Stuermer}, {Suchyta}, {Sypniewski},
  {Tarle}, {Thaler}, {Tighe}, {Tran}, {Tucker}, {Walker}, {Wang}, {Watson},
  {Weaverdyck}, {Wester}, {Woods}, {Yanny}, \& {DES
  Collaboration}}]{Flaugher2015}
{Flaugher}, B., {Diehl}, H.~T., {Honscheid}, K., {et~al.} 2015, \aj, 150, 150,
  \dodoi{10.1088/0004-6256/150/5/150}

\bibitem[{{Frenk} \& {White}(2012)}]{Frenk2012}
{Frenk}, C.~S., \& {White}, S.~D.~M. 2012, Annalen der Physik, 524, 507,
  \dodoi{10.1002/andp.201200212}

\bibitem[{{Gaia Collaboration} {et~al.}(2020){Gaia Collaboration}, {Brown},
  {Vallenari}, {Prusti}, {de Bruijne}, {Babusiaux}, \& {Biermann}}]{GaiaEDR3}
{Gaia Collaboration}, {Brown}, A.~G.~A., {Vallenari}, A., {et~al.} 2020, arXiv
  e-prints, arXiv:2012.01533.
\newblock \doarXiv{2012.01533}

\bibitem[{{Gaia Collaboration} {et~al.}(2016){Gaia Collaboration}, {Prusti},
  {de Bruijne}, {Brown}, {Vallenari}, {Babusiaux}, {Bailer-Jones}, {Bastian},
  {Biermann}, {Evans}, {Eyer}, {Jansen}, {Jordi}, {Klioner}, {Lammers},
  {Lindegren}, {Luri}, {Mignard}, {Milligan}, {Panem}, {Poinsignon},
  {Pourbaix}, {Randich}, {Sarri}, {Sartoretti}, {Siddiqui}, {Soubiran},
  {Valette}, {van Leeuwen}, {Walton}, {Aerts}, {Arenou}, {Cropper}, {Drimmel},
  {H{\o}g}, {Katz}, {Lattanzi}, {O'Mullane}, {Grebel}, {Holland}, {Huc},
  {Passot}, {Bramante}, {Cacciari}, {Casta{\~n}eda}, {Chaoul}, {Cheek}, {De
  Angeli}, {Fabricius}, {Guerra}, {Hern{\'a}ndez}, {Jean-Antoine-Piccolo},
  {Masana}, {Messineo}, {Mowlavi}, {Nienartowicz}, {Ord{\'o}{\~n}ez-Blanco},
  {Panuzzo}, {Portell}, {Richards}, {Riello}, {Seabroke}, {Tanga},
  {Th{\'e}venin}, {Torra}, {Els}, {Gracia-Abril}, {Comoretto},
  {Garcia-Reinaldos}, {Lock}, {Mercier}, {Altmann}, {Andrae}, {Astraatmadja},
  {Bellas-Velidis}, {Benson}, {Berthier}, {Blomme}, {Busso}, {Carry},
  {Cellino}, {Clementini}, {Cowell}, {Creevey}, {Cuypers}, {Davidson}, {De
  Ridder}, {de Torres}, {Delchambre}, {Dell'Oro}, {Ducourant}, {Fr{\'e}mat},
  {Garc{\'\i}a-Torres}, {Gosset}, {Halbwachs}, {Hambly}, {Harrison}, {Hauser},
  {Hestroffer}, {Hodgkin}, {Huckle}, {Hutton}, {Jasniewicz}, {Jordan},
  {Kontizas}, {Korn}, {Lanzafame}, {Manteiga}, {Moitinho}, {Muinonen},
  {Osinde}, {Pancino}, {Pauwels}, {Petit}, {Recio-Blanco}, {Robin}, {Sarro},
  {Siopis}, {Smith}, {Smith}, {Sozzetti}, {Thuillot}, {van Reeven}, {Viala},
  {Abbas}, {Abreu Aramburu}, {Accart}, {Aguado}, {Allan}, {Allasia},
  {Altavilla}, {{\'A}lvarez}, {Alves}, {Anderson}, {Andrei}, {Anglada Varela},
  {Antiche}, {Antoja}, {Ant{\'o}n}, {Arcay}, {Atzei}, {Ayache}, {Bach},
  {Baker}, {Balaguer-N{\'u}{\~n}ez}, {Barache}, {Barata}, {Barbier}, {Barblan},
  {Baroni}, {Barrado y Navascu{\'e}s}, {Barros}, {Barstow}, {Becciani},
  {Bellazzini}, {Bellei}, {Bello Garc{\'\i}a}, {Belokurov}, {Bendjoya},
  {Berihuete}, {Bianchi}, {Bienaym{\'e}}, {Billebaud}, {Blagorodnova},
  {Blanco-Cuaresma}, {Boch}, {Bombrun}, {Borrachero}, {Bouquillon}, {Bourda},
  {Bouy}, {Bragaglia}, {Breddels}, {Brouillet}, {Br{\"u}semeister},
  {Bucciarelli}, {Budnik}, {Burgess}, {Burgon}, {Burlacu}, {Busonero}, {Buzzi},
  {Caffau}, {Cambras}, {Campbell}, {Cancelliere}, {Cantat-Gaudin}, {Carlucci},
  {Carrasco}, {Castellani}, {Charlot}, {Charnas}, {Charvet}, {Chassat},
  {Chiavassa}, {Clotet}, {Cocozza}, {Collins}, {Collins}, {Costigan}, {Crifo},
  {Cross}, {Crosta}, {Crowley}, {Dafonte}, {Damerdji}, {Dapergolas}, {David},
  {David}, {De Cat}, {de Felice}, {de Laverny}, {De Luise}, {De March}, {de
  Martino}, {de Souza}, {Debosscher}, {del Pozo}, {Delbo}, {Delgado},
  {Delgado}, {di Marco}, {Di Matteo}, {Diakite}, {Distefano}, {Dolding}, {Dos
  Anjos}, {Drazinos}, {Dur{\'a}n}, {Dzigan}, {Ecale}, {Edvardsson}, {Enke},
  {Erdmann}, {Escolar}, {Espina}, {Evans}, {Eynard Bontemps}, {Fabre},
  {Fabrizio}, {Faigler}, {Falc{\~a}o}, {Farr{\`a}s Casas}, {Faye}, {Federici},
  {Fedorets}, {Fern{\'a}ndez-Hern{\'a}ndez}, {Fernique}, {Fienga}, {Figueras},
  {Filippi}, {Findeisen}, {Fonti}, {Fouesneau}, {Fraile}, {Fraser}, {Fuchs},
  {Furnell}, {Gai}, {Galleti}, {Galluccio}, {Garabato}, {Garc{\'\i}a-Sedano},
  {Gar{\'e}}, {Garofalo}, {Garralda}, {Gavras}, {Gerssen}, {Geyer}, {Gilmore},
  {Girona}, {Giuffrida}, {Gomes}, {Gonz{\'a}lez-Marcos},
  {Gonz{\'a}lez-N{\'u}{\~n}ez}, {Gonz{\'a}lez-Vidal}, {Granvik}, {Guerrier},
  {Guillout}, {Guiraud}, {G{\'u}rpide}, {Guti{\'e}rrez-S{\'a}nchez}, {Guy},
  {Haigron}, {Hatzidimitriou}, {Haywood}, {Heiter}, {Helmi}, {Hobbs},
  {Hofmann}, {Holl}, {Holland}, {Hunt}, {Hypki}, {Icardi}, {Irwin}, {Jevardat
  de Fombelle}, {Jofr{\'e}}, {Jonker}, {Jorissen}, {Julbe}, {Karampelas},
  {Kochoska}, {Kohley}, {Kolenberg}, {Kontizas}, {Koposov}, {Kordopatis},
  {Koubsky}, {Kowalczyk}, {Krone-Martins}, {Kudryashova}, {Kull}, {Bachchan},
  {Lacoste-Seris}, {Lanza}, {Lavigne}, {Le Poncin-Lafitte}, {Lebreton},
  {Lebzelter}, {Leccia}, {Leclerc}, {Lecoeur-Taibi}, {Lemaitre}, {Lenhardt},
  {Leroux}, {Liao}, {Licata}, {Lindstr{\o}m}, {Lister}, {Livanou}, {Lobel},
  {L{\"o}ffler}, {L{\'o}pez}, {Lopez-Lozano}, {Lorenz}, {Loureiro},
  {MacDonald}, {Magalh{\~a}es Fernandes}, {Managau}, {Mann}, {Mantelet},
  {Marchal}, {Marchant}, {Marconi}, {Marie}, {Marinoni}, {Marrese},
  {Marschalk{\'o}}, {Marshall}, {Mart{\'\i}n-Fleitas}, {Martino}, {Mary},
  {Matijevi{\v{c}}}, {Mazeh}, {McMillan}, {Messina}, {Mestre}, {Michalik},
  {Millar}, {Miranda}, {Molina}, {Molinaro}, {Molinaro}, {Moln{\'a}r},
  {Moniez}, {Montegriffo}, {Monteiro}, {Mor}, {Mora}, {Morbidelli}, {Morel},
  {Morgenthaler}, {Morley}, {Morris}, {Mulone}, {Muraveva}, {Musella},
  {Narbonne}, {Nelemans}, {Nicastro}, {Noval}, {Ord{\'e}novic},
  {Ordieres-Mer{\'e}}, {Osborne}, {Pagani}, {Pagano}, {Pailler}, {Palacin},
  {Palaversa}, {Parsons}, {Paulsen}, {Pecoraro}, {Pedrosa}, {Pentik{\"a}inen},
  {Pereira}, {Pichon}, {Piersimoni}, {Pineau}, {Plachy}, {Plum}, {Poujoulet},
  {Pr{\v{s}}a}, {Pulone}, {Ragaini}, {Rago}, {Rambaux}, {Ramos-Lerate},
  {Ranalli}, {Rauw}, {Read}, {Regibo}, {Renk}, {Reyl{\'e}}, {Ribeiro},
  {Rimoldini}, {Ripepi}, {Riva}, {Rixon}, {Roelens}, {Romero-G{\'o}mez},
  {Rowell}, {Royer}, {Rudolph}, {Ruiz-Dern}, {Sadowski}, {Sagrist{\`a}
  Sell{\'e}s}, {Sahlmann}, {Salgado}, {Salguero}, {Sarasso}, {Savietto},
  {Schnorhk}, {Schultheis}, {Sciacca}, {Segol}, {Segovia}, {Segransan},
  {Serpell}, {Shih}, {Smareglia}, {Smart}, {Smith}, {Solano}, {Solitro},
  {Sordo}, {Soria Nieto}, {Souchay}, {Spagna}, {Spoto}, {Stampa}, {Steele},
  {Steidelm{\"u}ller}, {Stephenson}, {Stoev}, {Suess}, {S{\"u}veges}, {Surdej},
  {Szabados}, {Szegedi-Elek}, {Tapiador}, {Taris}, {Tauran}, {Taylor},
  {Teixeira}, {Terrett}, {Tingley}, {Trager}, {Turon}, {Ulla}, {Utrilla},
  {Valentini}, {van Elteren}, {Van Hemelryck}, {van Leeuwen}, {Varadi},
  {Vecchiato}, {Veljanoski}, {Via}, {Vicente}, {Vogt}, {Voss}, {Votruba},
  {Voutsinas}, {Walmsley}, {Weiler}, {Weingrill}, {Werner}, {Wevers},
  {Whitehead}, {Wyrzykowski}, {Yoldas}, {{\v{Z}}erjal}, {Zucker}, {Zurbach},
  {Zwitter}, {Alecu}, {Allen}, {Allende Prieto}, {Amorim},
  {Anglada-Escud{\'e}}, {Arsenijevic}, {Azaz}, {Balm}, {Beck}, {Bernstein},
  {Bigot}, {Bijaoui}, {Blasco}, {Bonfigli}, {Bono}, {Boudreault}, {Bressan},
  {Brown}, {Brunet}, {Bunclark}, {Buonanno}, {Butkevich}, {Carret}, {Carrion},
  {Chemin}, {Ch{\'e}reau}, {Corcione}, {Darmigny}, {de Boer}, {de Teodoro}, {de
  Zeeuw}, {Delle Luche}, {Domingues}, {Dubath}, {Fodor}, {Fr{\'e}zouls},
  {Fries}, {Fustes}, {Fyfe}, {Gallardo}, {Gallegos}, {Gardiol}, {Gebran},
  {Gomboc}, {G{\'o}mez}, {Grux}, {Gueguen}, {Heyrovsky}, {Hoar}, {Iannicola},
  {Isasi Parache}, {Janotto}, {Joliet}, {Jonckheere}, {Keil}, {Kim},
  {Klagyivik}, {Klar}, {Knude}, {Kochukhov}, {Kolka}, {Kos}, {Kutka}, {Lainey},
  {LeBouquin}, {Liu}, {Loreggia}, {Makarov}, {Marseille}, {Martayan},
  {Martinez-Rubi}, {Massart}, {Meynadier}, {Mignot}, {Munari}, {Nguyen},
  {Nordlander}, {Ocvirk}, {O'Flaherty}, {Olias Sanz}, {Ortiz}, {Osorio},
  {Oszkiewicz}, {Ouzounis}, {Palmer}, {Park}, {Pasquato}, {Peltzer}, {Peralta},
  {P{\'e}turaud}, {Pieniluoma}, {Pigozzi}, {Poels}, {Prat}, {Prod'homme},
  {Raison}, {Rebordao}, {Risquez}, {Rocca-Volmerange}, {Rosen}, {Ruiz-Fuertes},
  {Russo}, {Sembay}, {Serraller Vizcaino}, {Short}, {Siebert}, {Silva},
  {Sinachopoulos}, {Slezak}, {Soffel}, {Sosnowska}, {Strai{\v{z}}ys}, {ter
  Linden}, {Terrell}, {Theil}, {Tiede}, {Troisi}, {Tsalmantza}, {Tur},
  {Vaccari}, {Vachier}, {Valles}, {Van Hamme}, {Veltz}, {Virtanen}, {Wallut},
  {Wichmann}, {Wilkinson}, {Ziaeepour}, \& {Zschocke}}]{GaiaMission}
{Gaia Collaboration}, {Prusti}, T., {de Bruijne}, J.~H.~J., {et~al.} 2016,
  \aap, 595, A1, \dodoi{10.1051/0004-6361/201629272}

\bibitem[{{Gaia Collaboration} {et~al.}(2018){Gaia Collaboration}, {Brown},
  {Vallenari}, {Prusti}, {de Bruijne}, {Babusiaux}, {Bailer-Jones}, {Biermann},
  {Evans}, {Eyer}, {Jansen}, {Jordi}, {Klioner}, {Lammers}, {Lindegren},
  {Luri}, {Mignard}, {Panem}, {Pourbaix}, {Randich}, {Sartoretti}, {Siddiqui},
  {Soubiran}, {van Leeuwen}, {Walton}, {Arenou}, {Bastian}, {Cropper},
  {Drimmel}, {Katz}, {Lattanzi}, {Bakker}, {Cacciari}, {Casta{\~n}eda},
  {Chaoul}, {Cheek}, {De Angeli}, {Fabricius}, {Guerra}, {Holl}, {Masana},
  {Messineo}, {Mowlavi}, {Nienartowicz}, {Panuzzo}, {Portell}, {Riello},
  {Seabroke}, {Tanga}, {Th{\'e}venin}, {Gracia-Abril}, {Comoretto},
  {Garcia-Reinaldos}, {Teyssier}, {Altmann}, {Andrae}, {Audard},
  {Bellas-Velidis}, {Benson}, {Berthier}, {Blomme}, {Burgess}, {Busso},
  {Carry}, {Cellino}, {Clementini}, {Clotet}, {Creevey}, {Davidson}, {De
  Ridder}, {Delchambre}, {Dell'Oro}, {Ducourant}, {Fern{\'a}ndez-
  Hern{\'a}ndez}, {Fouesneau}, {Fr{\'e}mat}, {Galluccio}, {Garc{\'\i}a-Torres},
  {Gonz{\'a}lez-N{\'u}{\~n}ez}, {Gonz{\'a}lez-Vidal}, {Gosset}, {Guy},
  {Halbwachs}, {Hambly}, {Harrison}, {Hern{\'a}ndez}, {Hestroffer}, {Hodgkin},
  {Hutton}, {Jasniewicz}, {Jean-Antoine-Piccolo}, {Jordan}, {Korn},
  {Krone-Martins}, {Lanzafame}, {Lebzelter}, {L{\"o}ffler}, {Manteiga},
  {Marrese}, {Mart{\'\i}n-Fleitas}, {Moitinho}, {Mora}, {Muinonen}, {Osinde},
  {Pancino}, {Pauwels}, {Petit}, {Recio-Blanco}, {Richards}, {Rimoldini},
  {Robin}, {Sarro}, {Siopis}, {Smith}, {Sozzetti}, {S{\"u}veges}, {Torra}, {van
  Reeven}, {Abbas}, {Abreu Aramburu}, {Accart}, {Aerts}, {Altavilla},
  {{\'A}lvarez}, {Alvarez}, {Alves}, {Anderson}, {Andrei}, {Anglada Varela},
  {Antiche}, {Antoja}, {Arcay}, {Astraatmadja}, {Bach}, {Baker},
  {Balaguer-N{\'u}{\~n}ez}, {Balm}, {Barache}, {Barata}, {Barbato}, {Barblan},
  {Barklem}, {Barrado}, {Barros}, {Barstow}, {Bartholom{\'e} Mu{\~n}oz},
  {Bassilana}, {Becciani}, {Bellazzini}, {Berihuete}, {Bertone}, {Bianchi},
  {Bienaym{\'e}}, {Blanco-Cuaresma}, {Boch}, {Boeche}, {Bombrun}, {Borrachero},
  {Bossini}, {Bouquillon}, {Bourda}, {Bragaglia}, {Bramante}, {Breddels},
  {Bressan}, {Brouillet}, {Br{\"u}semeister}, {Brugaletta}, {Bucciarelli},
  {Burlacu}, {Busonero}, {Butkevich}, {Buzzi}, {Caffau}, {Cancelliere},
  {Cannizzaro}, {Cantat-Gaudin}, {Carballo}, {Carlucci}, {Carrasco},
  {Casamiquela}, {Castellani}, {Castro-Ginard}, {Charlot}, {Chemin},
  {Chiavassa}, {Cocozza}, {Costigan}, {Cowell}, {Crifo}, {Crosta}, {Crowley},
  {Cuypers}, {Dafonte}, {Damerdji}, {Dapergolas}, {David}, {David}, {de
  Laverny}, {De Luise}, {De March}, {de Martino}, {de Souza}, {de Torres},
  {Debosscher}, {del Pozo}, {Delbo}, {Delgado}, {Delgado}, {Di Matteo},
  {Diakite}, {Diener}, {Distefano}, {Dolding}, {Drazinos}, {Dur{\'a}n},
  {Edvardsson}, {Enke}, {Eriksson}, {Esquej}, {Eynard Bontemps}, {Fabre},
  {Fabrizio}, {Faigler}, {Falc{\~a}o}, {Farr{\`a}s Casas}, {Federici},
  {Fedorets}, {Fernique}, {Figueras}, {Filippi}, {Findeisen}, {Fonti},
  {Fraile}, {Fraser}, {Fr{\'e}zouls}, {Gai}, {Galleti}, {Garabato},
  {Garc{\'\i}a-Sedano}, {Garofalo}, {Garralda}, {Gavel}, {Gavras}, {Gerssen},
  {Geyer}, {Giacobbe}, {Gilmore}, {Girona}, {Giuffrida}, {Glass}, {Gomes},
  {Granvik}, {Gueguen}, {Guerrier}, {Guiraud}, {Guti{\'e}rrez-S{\'a}nchez},
  {Haigron}, {Hatzidimitriou}, {Hauser}, {Haywood}, {Heiter}, {Helmi}, {Heu},
  {Hilger}, {Hobbs}, {Hofmann}, {Holland}, {Huckle}, {Hypki}, {Icardi},
  {Jan{\ss}en}, {Jevardat de Fombelle}, {Jonker}, {Juh{\'a}sz}, {Julbe},
  {Karampelas}, {Kewley}, {Klar}, {Kochoska}, {Kohley}, {Kolenberg},
  {Kontizas}, {Kontizas}, {Koposov}, {Kordopatis}, {Kostrzewa-Rutkowska},
  {Koubsky}, {Lambert}, {Lanza}, {Lasne}, {Lavigne}, {Le Fustec}, {Le
  Poncin-Lafitte}, {Lebreton}, {Leccia}, {Leclerc}, {Lecoeur-Taibi},
  {Lenhardt}, {Leroux}, {Liao}, {Licata}, {Lindstr{\o}m}, {Lister}, {Livanou},
  {Lobel}, {L{\'o}pez}, {Managau}, {Mann}, {Mantelet}, {Marchal}, {Marchant},
  {Marconi}, {Marinoni}, {Marschalk{\'o}}, {Marshall}, {Martino}, {Marton},
  {Mary}, {Massari}, {Matijevi{\v{c}}}, {Mazeh}, {McMillan}, {Messina},
  {Michalik}, {Millar}, {Molina}, {Molinaro}, {Moln{\'a}r}, {Montegriffo},
  {Mor}, {Morbidelli}, {Morel}, {Morris}, {Mulone}, {Muraveva}, {Musella},
  {Nelemans}, {Nicastro}, {Noval}, {O'Mullane}, {Ord{\'e}novic},
  {Ord{\'o}{\~n}ez-Blanco}, {Osborne}, {Pagani}, {Pagano}, {Pailler},
  {Palacin}, {Palaversa}, {Panahi}, {Pawlak}, {Piersimoni}, {Pineau}, {Plachy},
  {Plum}, {Poggio}, {Poujoulet}, {Pr{\v{s}}a}, {Pulone}, {Racero}, {Ragaini},
  {Rambaux}, {Ramos-Lerate}, {Regibo}, {Reyl{\'e}}, {Riclet}, {Ripepi}, {Riva},
  {Rivard}, {Rixon}, {Roegiers}, {Roelens}, {Romero-G{\'o}mez}, {Rowell},
  {Royer}, {Ruiz-Dern}, {Sadowski}, {Sagrist{\`a} Sell{\'e}s}, {Sahlmann},
  {Salgado}, {Salguero}, {Sanna}, {Santana- Ros}, {Sarasso}, {Savietto},
  {Schultheis}, {Sciacca}, {Segol}, {Segovia}, {S{\'e}gransan}, {Shih},
  {Siltala}, {Silva}, {Smart}, {Smith}, {Solano}, {Solitro}, {Sordo}, {Soria
  Nieto}, {Souchay}, {Spagna}, {Spoto}, {Stampa}, {Steele},
  {Steidelm{\"u}ller}, {Stephenson}, {Stoev}, {Suess}, {Surdej}, {Szabados},
  {Szegedi-Elek}, {Tapiador}, {Taris}, {Tauran}, {Taylor}, {Teixeira},
  {Terrett}, {Teyssandier}, {Thuillot}, {Titarenko}, {Torra Clotet}, {Turon},
  {Ulla}, {Utrilla}, {Uzzi}, {Vaillant}, {Valentini}, {Valette}, {van Elteren},
  {Van Hemelryck}, {van Leeuwen}, {Vaschetto}, {Vecchiato}, {Veljanoski},
  {Viala}, {Vicente}, {Vogt}, {von Essen}, {Voss}, {Votruba}, {Voutsinas},
  {Walmsley}, {Weiler}, {Wertz}, {Wevers}, {Wyrzykowski}, {Yoldas},
  {{\v{Z}}erjal}, {Ziaeepour}, {Zorec}, {Zschocke}, {Zucker}, {Zurbach}, \&
  {Zwitter}}]{GaiaDR2}
{Gaia Collaboration}, {Brown}, A.~G.~A., {Vallenari}, A., {et~al.} 2018, \aap,
  616, A1, \dodoi{10.1051/0004-6361/201833051}

\bibitem[{{Hidalgo} {et~al.}(2018){Hidalgo}, {Pietrinferni}, {Cassisi},
  {Salaris}, {Mucciarelli}, {Savino}, {Aparicio}, {Silva Aguirre}, \&
  {Verma}}]{Hidalgo2018}
{Hidalgo}, S.~L., {Pietrinferni}, A., {Cassisi}, S., {et~al.} 2018, \apj, 856,
  125, \dodoi{10.3847/1538-4357/aab158}

\bibitem[{{Holl} {et~al.}(2018){Holl}, {Audard}, {Nienartowicz}, {Jevardat de
  Fombelle}, {Marchal}, {Mowlavi}, {Clementini}, {De Ridder}, {Evans}, {Guy},
  {Lanzafame}, {Lebzelter}, {Rimoldini}, {Roelens}, {Zucker}, {Distefano},
  {Garofalo}, {Lecoeur-Ta{\"\i}bi}, {Lopez}, {Molinaro}, {Muraveva}, {Panahi},
  {Regibo}, {Ripepi}, {Sarro}, {Aerts}, {Anderson}, {Charnas}, {Barblan},
  {Blanco-Cuaresma}, {Busso}, {Cuypers}, {De Angeli}, {Glass}, {Grenon},
  {Juh{\'a}sz}, {Kochoska}, {Koubsky}, {Lanza}, {Leccia}, {Lorenz}, {Marconi},
  {Marschalk{\'o}}, {Mazeh}, {Messina}, {Mignard}, {Moitinho}, {Moln{\'a}r},
  {Morgenthaler}, {Musella}, {Ordenovic}, {Ord{\'o}{\~n}ez}, {Pagano},
  {Palaversa}, {Pawlak}, {Plachy}, {Pr{\v{s}}a}, {Riello}, {S{\"u}veges},
  {Szabados}, {Szegedi-Elek}, {Votruba}, \& {Eyer}}]{Holl2018}
{Holl}, B., {Audard}, M., {Nienartowicz}, K., {et~al.} 2018, \aap, 618, A30,
  \dodoi{10.1051/0004-6361/201832892}

\bibitem[{{Horne} \& {Baliunas}(1986)}]{Horne1986}
{Horne}, J.~H., \& {Baliunas}, S.~L. 1986, \apj, 302, 757,
  \dodoi{10.1086/164037}

\bibitem[{Hunter(2007)}]{Matplotlib}
Hunter, J.~D. 2007, Computing in Science \& Engineering, 9, 90,
  \dodoi{10.1109/MCSE.2007.55}

\bibitem[{{Ivezi{\'c}} {et~al.}(2019){Ivezi{\'c}}, {Kahn}, {Tyson}, {Abel},
  {Acosta}, {Allsman}, {Alonso}, {AlSayyad}, {Anderson}, {Andrew}, {Angel},
  {Angeli}, {Ansari}, {Antilogus}, {Araujo}, {Armstrong}, {Arndt}, {Astier},
  {Aubourg}, {Auza}, {Axelrod}, {Bard}, {Barr}, {Barrau}, {Bartlett}, {Bauer},
  {Bauman}, {Baumont}, {Bechtol}, {Bechtol}, {Becker}, {Becla}, {Beldica},
  {Bellavia}, {Bianco}, {Biswas}, {Blanc}, {Blazek}, {Blandford}, {Bloom},
  {Bogart}, {Bond}, {Booth}, {Borgland}, {Borne}, {Bosch}, {Boutigny},
  {Brackett}, {Bradshaw}, {Brandt}, {Brown}, {Bullock}, {Burchat}, {Burke},
  {Cagnoli}, {Calabrese}, {Callahan}, {Callen}, {Carlin}, {Carlson},
  {Chandrasekharan}, {Charles-Emerson}, {Chesley}, {Cheu}, {Chiang}, {Chiang},
  {Chirino}, {Chow}, {Ciardi}, {Claver}, {Cohen-Tanugi}, {Cockrum}, {Coles},
  {Connolly}, {Cook}, {Cooray}, {Covey}, {Cribbs}, {Cui}, {Cutri}, {Daly},
  {Daniel}, {Daruich}, {Daubard}, {Daues}, {Dawson}, {Delgado}, {Dellapenna},
  {de Peyster}, {de Val-Borro}, {Digel}, {Doherty}, {Dubois},
  {Dubois-Felsmann}, {Durech}, {Economou}, {Eifler}, {Eracleous}, {Emmons},
  {Fausti Neto}, {Ferguson}, {Figueroa}, {Fisher-Levine}, {Focke}, {Foss},
  {Frank}, {Freemon}, {Gangler}, {Gawiser}, {Geary}, {Gee}, {Geha}, {Gessner},
  {Gibson}, {Gilmore}, {Glanzman}, {Glick}, {Goldina}, {Goldstein}, {Goodenow},
  {Graham}, {Gressler}, {Gris}, {Guy}, {Guyonnet}, {Haller}, {Harris},
  {Hascall}, {Haupt}, {Hernandez}, {Herrmann}, {Hileman}, {Hoblitt}, {Hodgson},
  {Hogan}, {Howard}, {Huang}, {Huffer}, {Ingraham}, {Innes}, {Jacoby}, {Jain},
  {Jammes}, {Jee}, {Jenness}, {Jernigan}, {Jevremovi{\'c}}, {Johns}, {Johnson},
  {Johnson}, {Jones}, {Juramy-Gilles}, {Juri{\'c}}, {Kalirai}, {Kallivayalil},
  {Kalmbach}, {Kantor}, {Karst}, {Kasliwal}, {Kelly}, {Kessler}, {Kinnison},
  {Kirkby}, {Knox}, {Kotov}, {Krabbendam}, {Krughoff}, {Kub{\'a}nek},
  {Kuczewski}, {Kulkarni}, {Ku}, {Kurita}, {Lage}, {Lambert}, {Lange},
  {Langton}, {Le Guillou}, {Levine}, {Liang}, {Lim}, {Lintott}, {Long},
  {Lopez}, {Lotz}, {Lupton}, {Lust}, {MacArthur}, {Mahabal}, {Mandelbaum},
  {Markiewicz}, {Marsh}, {Marshall}, {Marshall}, {May}, {McKercher}, {McQueen},
  {Meyers}, {Migliore}, {Miller}, {Mills}, {Miraval}, {Moeyens}, {Moolekamp},
  {Monet}, {Moniez}, {Monkewitz}, {Montgomery}, {Morrison}, {Mueller},
  {Muller}, {Mu{\~n}oz Arancibia}, {Neill}, {Newbry}, {Nief}, {Nomerotski},
  {Nordby}, {O'Connor}, {Oliver}, {Olivier}, {Olsen}, {O'Mullane}, {Ortiz},
  {Osier}, {Owen}, {Pain}, {Palecek}, {Parejko}, {Parsons}, {Pease},
  {Peterson}, {Peterson}, {Petravick}, {Libby Petrick}, {Petry},
  {Pierfederici}, {Pietrowicz}, {Pike}, {Pinto}, {Plante}, {Plate}, {Plutchak},
  {Price}, {Prouza}, {Radeka}, {Rajagopal}, {Rasmussen}, {Regnault}, {Reil},
  {Reiss}, {Reuter}, {Ridgway}, {Riot}, {Ritz}, {Robinson}, {Roby}, {Roodman},
  {Rosing}, {Roucelle}, {Rumore}, {Russo}, {Saha}, {Sassolas}, {Schalk},
  {Schellart}, {Schindler}, {Schmidt}, {Schneider}, {Schneider}, {Schoening},
  {Schumacher}, {Schwamb}, {Sebag}, {Selvy}, {Sembroski}, {Seppala}, {Serio},
  {Serrano}, {Shaw}, {Shipsey}, {Sick}, {Silvestri}, {Slater}, {Smith},
  {Smith}, {Sobhani}, {Soldahl}, {Storrie-Lombardi}, {Stover}, {Strauss},
  {Street}, {Stubbs}, {Sullivan}, {Sweeney}, {Swinbank}, {Szalay}, {Takacs},
  {Tether}, {Thaler}, {Thayer}, {Thomas}, {Thornton}, {Thukral}, {Tice},
  {Trilling}, {Turri}, {Van Berg}, {Vanden Berk}, {Vetter}, {Virieux},
  {Vucina}, {Wahl}, {Walkowicz}, {Walsh}, {Walter}, {Wang}, {Wang}, {Warner},
  {Wiecha}, {Willman}, {Winters}, {Wittman}, {Wolff}, {Wood-Vasey}, {Wu},
  {Xin}, {Yoachim}, \& {Zhan}}]{Ivezic2019}
{Ivezi{\'c}}, {\v{Z}}., {Kahn}, S.~M., {Tyson}, J.~A., {et~al.} 2019, \apj,
  873, 111, \dodoi{10.3847/1538-4357/ab042c}

\bibitem[{{Ji} {et~al.}(2019){Ji}, {Simon}, {Frebel}, {Venn}, \&
  {Hansen}}]{Ji2019}
{Ji}, A.~P., {Simon}, J.~D., {Frebel}, A., {Venn}, K.~A., \& {Hansen}, T.~T.
  2019, \apj, 870, 83, \dodoi{10.3847/1538-4357/aaf3bb}

\bibitem[{{Kunder} {et~al.}(2013){Kunder}, {Stetson}, {Cassisi}, {Layden},
  {Bono}, {Catelan}, {Walker}, {Paredes Alvarez}, {Clem}, {Matsunaga},
  {Salaris}, {Lee}, \& {Chaboyer}}]{Kunder2013}
{Kunder}, A., {Stetson}, P.~B., {Cassisi}, S., {et~al.} 2013, \aj, 146, 119,
  \dodoi{10.1088/0004-6256/146/5/119}

\bibitem[{{Layden}(1998)}]{Layden1998}
{Layden}, A.~C. 1998, \aj, 115, 193, \dodoi{10.1086/300195}

\bibitem[{{Li} {et~al.}(2019){Li}, {Koposov}, {Zucker}, {Lewis}, {Kuehn},
  {Simpson}, {Ji}, {Shipp}, {Mao}, {Geha}, {Pace}, {Mackey}, {Allam}, {Tucker},
  {Da Costa}, {Erkal}, {Simon}, {Mould}, {Martell}, {Wan}, {De Silva},
  {Bechtol}, {Balbinot}, {Belokurov}, {Bland-Hawthorn}, {Casey}, {Cullinane},
  {Drlica-Wagner}, {Sharma}, {Vivas}, {Wechsler}, {Yanny}, \& {S5
  Collaboration}}]{Li2019}
{Li}, T.~S., {Koposov}, S.~E., {Zucker}, D.~B., {et~al.} 2019, \mnras, 490,
  3508, \dodoi{10.1093/mnras/stz2731}

\bibitem[{{Marconi} {et~al.}(2015){Marconi}, {Coppola}, {Bono}, {Braga},
  {Pietrinferni}, {Buonanno}, {Castellani}, {Musella}, {Ripepi}, \&
  {Stellingwerf}}]{Marconi2015}
{Marconi}, M., {Coppola}, G., {Bono}, G., {et~al.} 2015, \apj, 808, 50,
  \dodoi{10.1088/0004-637X/808/1/50}

\bibitem[{{Mart{\'\i}nez-V{\'a}zquez}
  {et~al.}(2021){Mart{\'\i}nez-V{\'a}zquez}, {Salinas}, \&
  {Vivas}}]{MartinezVazquez2021}
{Mart{\'\i}nez-V{\'a}zquez}, C.~E., {Salinas}, R., \& {Vivas}, A.~K. 2021, \aj,
  161, 120, \dodoi{10.3847/1538-3881/abd55e}

\bibitem[{{Mart{\'{\i}}nez-V{\'a}zquez}
  {et~al.}(2015){Mart{\'{\i}}nez-V{\'a}zquez}, {Monelli}, {Bono}, {Stetson},
  {Ferraro}, {Bernard}, {Gallart}, {Fiorentino}, {Iannicola}, \&
  {Udalski}}]{MartinezVazquez2015}
{Mart{\'{\i}}nez-V{\'a}zquez}, C.~E., {Monelli}, M., {Bono}, G., {et~al.} 2015,
  \mnras, 454, 1509, \dodoi{10.1093/mnras/stv2014}

\bibitem[{{Mart{\'{\i}}nez-V{\'a}zquez}
  {et~al.}(2016){Mart{\'{\i}}nez-V{\'a}zquez}, {Stetson}, {Monelli}, {Bernard},
  {Fiorentino}, {Gallart}, {Bono}, {Cassisi}, {Dall'Ora}, {Ferraro},
  {Iannicola}, \& {Walker}}]{MartinezVazquez2016b}
{Mart{\'{\i}}nez-V{\'a}zquez}, C.~E., {Stetson}, P.~B., {Monelli}, M., {et~al.}
  2016, \mnras, 462, 4349, \dodoi{10.1093/mnras/stw1895}

\bibitem[{{Mart{\'{\i}}nez-V{\'a}zquez}
  {et~al.}(2017){Mart{\'{\i}}nez-V{\'a}zquez}, {Monelli}, {Bernard}, {Gallart},
  {Stetson}, {Skillman}, {Bono}, {Cassisi}, {Fiorentino}, {McQuinn}, {Cole},
  {McConnachie}, {Martin}, {Dolphin}, {Boylan-Kolchin}, {Aparicio}, {Hidalgo},
  \& {Weisz}}]{MartinezVazquez2017}
{Mart{\'{\i}}nez-V{\'a}zquez}, C.~E., {Monelli}, M., {Bernard}, E.~J., {et~al.}
  2017, \apj, 850, 137, \dodoi{10.3847/1538-4357/aa9381}

\bibitem[{{Mart{\'\i}nez-V{\'a}zquez}
  {et~al.}(2019){Mart{\'\i}nez-V{\'a}zquez}, {Vivas}, {Gurevich}, {Walker},
  {McCarthy}, {Pace}, {Stringer}, {Santiago}, {Hounsell}, {Macri}, {Li},
  {Bechtol}, {Riley}, {Kim}, {Simon}, {Drlica-Wagner}, {Nadler}, {Marshall},
  {Annis}, {Avila}, {Bertin}, {Brooks}, {Buckley-Geer}, {Burke}, {Carnero
  Rosell}, {Carrasco Kind}, {da Costa}, {De Vicente}, {Desai}, {Diehl}, {Doel},
  {Everett}, {Frieman}, {Garc{\'\i}a-Bellido}, {Gaztanaga}, {Gruen}, {Gruendl},
  {Gschwend}, {Gutierrez}, {Hollowood}, {Honscheid}, {James}, {Kuehn},
  {Kuropatkin}, {Maia}, {Menanteau}, {Miller}, {Miquel}, {Paz-Chinch{\'o}n},
  {Plazas}, {Sanchez}, {Scarpine}, {Serrano}, {Sevilla-Noarbe}, {Smith},
  {Soares-Santos}, {Sobreira}, {Swanson}, {Tarle}, {Vikram}, \& {DES
  Collaboration}}]{MartinezVazquez2019}
{Mart{\'\i}nez-V{\'a}zquez}, C.~E., {Vivas}, A.~K., {Gurevich}, M., {et~al.}
  2019, \mnras, 490, 2183, \dodoi{10.1093/mnras/stz2609}

\bibitem[{{Mau} {et~al.}(2020){Mau}, {Cerny}, {Pace}, {Choi}, {Drlica-Wagner},
  {Santana-Silva}, {Riley}, {Erkal}, {Stringfellow}, {Adam{\'o}w}, {Carlin},
  {Gruendl}, {Hernandez-Lang}, {Kuropatkin}, {Li}, {Mart{\'\i}nez-V{\'a}zquez},
  {Morganson}, {Mutlu-Pakdil}, {Neilsen}, {Nidever}, {Olsen}, {Sand},
  {Tollerud}, {Tucker}, {Yanny}, {Zenteno}, {Allam}, {Barkhouse}, {Bechtol},
  {Bell}, {Balaji}, {Crnojevi{\'c}}, {Esteves}, {Ferguson}, {Gallart},
  {Hughes}, {James}, {Jethwa}, {Johnson}, {Kuehn}, {Majewski}, {Mao},
  {Massana}, {McNanna}, {Monachesi}, {Nadler}, {No{\"e}l}, {Palmese},
  {Paz-Chinchon}, {Pieres}, {Sanchez}, {Shipp}, {Simon}, {Soares-Santos},
  {Tavangar}, {van der Marel}, {Vivas}, {Walker}, \& {Wechsler}}]{Mau2020}
{Mau}, S., {Cerny}, W., {Pace}, A.~B., {et~al.} 2020, \apj, 890, 136,
  \dodoi{10.3847/1538-4357/ab6c67}

\bibitem[{{McConnachie} \& {Venn}(2020)}]{McConnachie2020}
{McConnachie}, A.~W., \& {Venn}, K.~A. 2020, Research Notes of the American
  Astronomical Society, 4, 229, \dodoi{10.3847/2515-5172/abd18b}

\bibitem[{{Medina} {et~al.}(2018){Medina}, {Mu{\~n}oz}, {Vivas}, {Carlin},
  {F{\"o}rster}, {Mart{\'{\i}}nez}, {Galbany}, {Gonz{\'a}lez-Gait{\'a}n},
  {Hamuy}, {de Jaeger}, {Maureira}, \& {San Mart{\'{\i}}n}}]{Medina2018}
{Medina}, G.~E., {Mu{\~n}oz}, R.~R., {Vivas}, A.~K., {et~al.} 2018, \apj, 855,
  43, \dodoi{10.3847/1538-4357/aaad02}

\bibitem[{{Morganson} {et~al.}(2018){Morganson}, {Gruendl}, {Menanteau},
  {Carrasco Kind}, {Chen}, {Daues}, {Drlica-Wagner}, {Friedel}, {Gower},
  {Johnson}, {Johnson}, {Kessler}, {Paz-Chinch{\'o}n}, {Petravick}, {Pond},
  {Yanny}, {Allam}, {Armstrong}, {Barkhouse}, {Bechtol}, {Benoit-L{\'e}vy},
  {Bernstein}, {Bertin}, {Buckley-Geer}, {Covarrubias}, {Desai}, {Diehl},
  {Goldstein}, {Gruen}, {Li}, {Lin}, {Marriner}, {Mohr}, {Neilsen}, {Ngeow},
  {Paech}, {Rykoff}, {Sako}, {Sevilla-Noarbe}, {Sheldon}, {Sobreira}, {Tucker},
  {Wester}, \& {DES Collaboration}}]{Morganson:2018}
{Morganson}, E., {Gruendl}, R.~A., {Menanteau}, F., {et~al.} 2018, \pasp, 130,
  074501, \dodoi{10.1088/1538-3873/aab4ef}

\bibitem[{{Neeley} {et~al.}(2015){Neeley}, {Marengo}, {Bono}, {Braga},
  {Dall'Ora}, {Stetson}, {Ferraro}, {Freedman}, {Iannicola}, {Madore},
  {Matsunaga}, {Monson}, {Persson}, {Scowcroft}, \& {Seibert}}]{Neeley2015}
{Neeley}, J.~R., {Marengo}, M., {Bono}, G., {et~al.} 2015, ArXiv e-prints.
\newblock \doarXiv{1505.07858}

\bibitem[{{Oosterhoff}(1939)}]{Oosterhoff1939}
{Oosterhoff}, P.~T. 1939, The Observatory, 62, 104

\bibitem[{{Oosterhoff}(1944)}]{Oosterhoff1944}
---. 1944, \bain, 10, 55

\bibitem[{{Pace} \& {Li}(2019)}]{Pace2019}
{Pace}, A.~B., \& {Li}, T.~S. 2019, \apj, 875, 77,
  \dodoi{10.3847/1538-4357/ab0aee}

\bibitem[{Paterno(2004)}]{Paterno2004}
Paterno, M. 2004, \dodoi{10.2172/15017262}

\bibitem[{{Pritzl} {et~al.}(2005){Pritzl}, {Venn}, \&
  {Irwin}}]{Pritzl2005:abundance}
{Pritzl}, B.~J., {Venn}, K.~A., \& {Irwin}, M. 2005, \aj, 130, 2140,
  \dodoi{10.1086/432911}

\bibitem[{{Salaris} \& {Cassisi}(2005)}]{Salaris2005}
{Salaris}, M., \& {Cassisi}, S. 2005, {Evolution of Stars and Stellar
  Populations}

\bibitem[{{Savino} {et~al.}(2020){Savino}, {Koch}, {Prudil}, {Kunder}, \&
  {Smolec}}]{Savino2020}
{Savino}, A., {Koch}, A., {Prudil}, Z., {Kunder}, A., \& {Smolec}, R. 2020,
  \aap, 641, A96, \dodoi{10.1051/0004-6361/202038305}

\bibitem[{{Schlafly} \& {Finkbeiner}(2011)}]{Schlafly:2011}
{Schlafly}, E.~F., \& {Finkbeiner}, D.~P. 2011, \apj, 737, 103,
  \dodoi{10.1088/0004-637X/737/2/103}

\bibitem[{{Schlegel} {et~al.}(1998){Schlegel}, {Finkbeiner}, \&
  {Davis}}]{Schlegel:1998}
{Schlegel}, D.~J., {Finkbeiner}, D.~P., \& {Davis}, M. 1998, \apj, 500, 525,
  \dodoi{10.1086/305772}

\bibitem[{{Searle} \& {Zinn}(1978)}]{Searle1978}
{Searle}, L., \& {Zinn}, R. 1978, \apj, 225, 357, \dodoi{10.1086/156499}

\bibitem[{{Simon}(2019)}]{Simon2019}
{Simon}, J.~D. 2019, arXiv e-prints, arXiv:1901.05465.
\newblock \doarXiv{1901.05465}

\bibitem[{{Smith}(1995)}]{Smith1995}
{Smith}, H.~A. 1995, Cambridge Astrophysics Series, 27

\bibitem[{Sokolovsky {et~al.}(2016)Sokolovsky, Gavras, Karampelas, Antipin,
  Bellas-Velidis, Benni, Bonanos, Burdanov, Derlopa, Hatzidimitriou,
  Khokhryakova, Kolesnikova, Korotkiy, Lapukhin, Moretti, Popov, Pouliasis,
  Samus, Spetsieri, Veselkov, Volkov, Yang, \& Zubareva}]{Sok:2016}
Sokolovsky, K.~V., Gavras, P., Karampelas, A., {et~al.} 2016, Monthly Notices
  of the Royal Astronomical Society, 464, 274, \dodoi{10.1093/mnras/stw2262}

\bibitem[{{Stetson} {et~al.}(2014){Stetson}, {Fiorentino}, {Bono}, {Bernard},
  {Monelli}, {Iannicola}, {Gallart}, \& {Ferraro}}]{Stetson2014}
{Stetson}, P.~B., {Fiorentino}, G., {Bono}, G., {et~al.} 2014, \pasp, 126, 616,
  \dodoi{10.1086/677352}

\bibitem[{{Tolstoy} {et~al.}(2009){Tolstoy}, {Hill}, \& {Tosi}}]{Tolstoy2009}
{Tolstoy}, E., {Hill}, V., \& {Tosi}, M. 2009, \araa, 47, 371,
  \dodoi{10.1146/annurev-astro-082708-101650}

\bibitem[{Tonry {et~al.}(2018)Tonry, Denneau, Flewelling, Heinze, Onken,
  Smartt, Stalder, Weiland, \& Wolf}]{Tonry:2018}
Tonry, J.~L., Denneau, L., Flewelling, H., {et~al.} 2018, The Astrophysical
  Journal, 867, 105, \dodoi{10.3847/1538-4357/aae386}

\bibitem[{{Vivas} {et~al.}(2019){Vivas}, {Alonso-Garc{\'\i}a}, {Mateo},
  {Walker}, \& {Howard}}]{Vivas2019}
{Vivas}, A.~K., {Alonso-Garc{\'\i}a}, J., {Mateo}, M., {Walker}, A., \&
  {Howard}, B. 2019, \aj, 157, 35, \dodoi{10.3847/1538-3881/aaf4f3}

\bibitem[{{Vivas} {et~al.}(2020){Vivas}, {Mart{\'\i}nez-V{\'a}zquez}, \&
  {Walker}}]{Vivas2020b}
{Vivas}, A.~K., {Mart{\'\i}nez-V{\'a}zquez}, C., \& {Walker}, A.~R. 2020,
  \apjs, 247, 35, \dodoi{10.3847/1538-4365/ab67c0}

\bibitem[{{Vivas} {et~al.}(2016){Vivas}, {Olsen}, {Blum}, {Nidever}, {Walker},
  {Martin}, {Besla}, {Gallart}, {van der Marel}, {Majewski}, {Kaleida},
  {Mu{\~n}oz}, {Saha}, {Conn}, \& {Jin}}]{Vivas2016a}
{Vivas}, A.~K., {Olsen}, K., {Blum}, R., {et~al.} 2016, \aj, 151, 118,
  \dodoi{10.3847/0004-6256/151/5/118}

\bibitem[{{Walker}(1989)}]{Walker1989}
{Walker}, A.~R. 1989, \pasp, 101, 570, \dodoi{10.1086/132470}

\bibitem[{{White} \& {Frenk}(1991)}]{White1991}
{White}, S.~D.~M., \& {Frenk}, C.~S. 1991, \apj, 379, 52,
  \dodoi{10.1086/170483}

\bibitem[{{Willman} {et~al.}(2005{\natexlab{a}}){Willman}, {Blanton}, {West},
  {Dalcanton}, {Hogg}, {Schneider}, {Wherry}, {Yanny}, \&
  {Brinkmann}}]{Willman2005a}
{Willman}, B., {Blanton}, M.~R., {West}, A.~A., {et~al.} 2005{\natexlab{a}},
  \aj, 129, 2692, \dodoi{10.1086/430214}

\bibitem[{{Willman} {et~al.}(2005{\natexlab{b}}){Willman}, {Dalcanton},
  {Martinez-Delgado}, {West}, {Blanton}, {Hogg}, {Barentine}, {Brewington},
  {Harvanek}, {Kleinman}, {Krzesinski}, {Long}, {Neilsen}, {Nitta}, \&
  {Snedden}}]{Willman2005b}
{Willman}, B., {Dalcanton}, J.~J., {Martinez-Delgado}, D., {et~al.}
  2005{\natexlab{b}}, \apjl, 626, L85, \dodoi{10.1086/431760}

\bibitem[{{Wolf} {et~al.}(2018){Wolf}, {Onken}, {Luvaul}, {Schmidt}, {Bessell},
  {Chang}, {Da Costa}, {Mackey}, {Martin-Jones}, {Murphy}, {Preston}, {Scalzo},
  {Shao}, {Smillie}, {Tisserand}, {White}, \& {Yuan}}]{Wolf:2018}
{Wolf}, C., {Onken}, C.~A., {Luvaul}, L.~C., {et~al.} 2018, \pasa, 35, e010,
  \dodoi{10.1017/pasa.2018.5}

\bibitem[{{Wolf} {et~al.}(2010){Wolf}, {Martinez}, {Bullock}, {Kaplinghat},
  {Geha}, {Mu{\~n}oz}, {Simon}, \& {Avedo}}]{Wolf2010}
{Wolf}, J., {Martinez}, G.~D., {Bullock}, J.~S., {et~al.} 2010, \mnras, 406,
  1220, \dodoi{10.1111/j.1365-2966.2010.16753.x}

\bibitem[{{Zinn} {et~al.}(2014){Zinn}, {Horowitz}, {Vivas}, {Baltay}, {Ellman},
  {Hadjiyska}, {Rabinowitz}, \& {Miller}}]{Zinn2014}
{Zinn}, R., {Horowitz}, B., {Vivas}, A.~K., {et~al.} 2014, \apj, 781, 22,
  \dodoi{10.1088/0004-637X/781/1/22}

\end{thebibliography}
